# Order and Chaos near Equilibrium Points in the Potential of Rotating Highly Irregular-shaped Celestial Bodies

Yu Jiang[1, 2], Hexi Baoyin[2], Xianyu Wang[2], Yang Yu[3], Hengnian Li[1], Chao Peng[4], Zhibin Zhang[1]

1. State Key Laboratory of Astronautic Dynamics, Xi'an Satellite Control Center, Xi'an 710043, China
2. School of Aerospace Engineering, Tsinghua University, Beijing 100084, China
3. Lagrange Laboratory, University of Nice Sophia Antipolis, CNRS, Observatoire de la Côte d'Azur, C.S. 34229, 06304 Nice Cedex 4, France
4. Technology and Engineering Center for Space Utilization, Chinese Academy of Sciences, Beijing 100094, China
Y. Jiang (✉) e-mail: jiangyu_xian_china@163.com (corresponding author)

**Abstract.** The order and chaos of the motion near equilibrium points in the potential of a rotating highly irregular-shaped celestial body are investigated from point of view of the dynamical system theory. The positions of the non-degenerate equilibrium points vary continuously when the parameter changes. The topological structures in the vicinity of equilibrium points are classified into several different cases. Bifurcations at equilibrium points and the topological transfers between different cases for equilibrium points are also discussed. The conclusions can be applied to all kinds of rotating celestial bodies, simple-shaped or highly irregular-shaped, including asteroids, comets, planets and satellites of planets to help one to understand the dynamical behaviors around them. Applications to asteroids 216 Kleopatra, 2063 Bacchus, and 25143 Itokawa are significant and interesting: eigenvalues affiliated to the equilibrium points for the asteroid 216 Kleopatra move and always belong to the same topological cases; while eigenvalues affiliated to two different equilibrium points for the asteroid 2063 Bacchus and 25143 Itokawa move through the resonant cases of equilibrium points, and the collision of eigenvalues in the complex plane occurs. Poincaré sections in the potential of the asteroid 216 Kleopatra show the chaos behaviors of the orbits in large scale.

# 1. Introduction

To the author's opinion, order and chaos are the pair of important and basic behaviors of the dynamical systems of natural and social worlds. Because of some space missions to asteroids and comets, people have an increasing interest of stability, bifurcations, chaos, and resonance of motion in the potential of non-spherical shaped celestial body. Dynamical behavior near some simple-shaped bodies or in some special potential fields have been discussed, these simple-shaped bodies and special potential fields includes a logarithmic gravity field [1], a straight segment [2-8], a circular ring [9-11], a rotating mass dipole[12], a homogeneous annulus disk [13-15], a triangular plate and a square plate [16], a homogeneous cube [17-20] and a dumbbell-shaped body [21].

Considering highly irregular-shaped minor bodies in the solar systems, objects which have been analyzed includes: Phobos [22]; asteroids 4 Vesta [23], 216 Kleopatra [24-27], 433 Eros [28-30], 1580 Betulia [30], 1620 Geographos [27,31], 4179 Toutatis [30, 32], 4769 Castalia [32,33-35], 6489 Golevka [31,36], 25143 Itokawa [37-38]; as well as comets 1/P Halley[36] and 67/P CG [30].

The existence, number and stability of equilibrium points are studied for some simple-shaped bodies in details. Elipe and Riaguas [1] found 4 equilibrium points and discussed their linear stability in the gravity field of a logarithm and a massive finite segment. Mondelo et al. [23] found 4 equilibrium points and gave their coordinates and stability near asteroid 4 Vesta, two of them are stable, while the other two unstable. Yu and Baoyin [24] found 4 equilibrium points and gave their coordinates, eigenvalues and stability near asteroid 216 Kleopatra. Scheeres [30] found 6 equilibrium points near asteroid 1580 Betulia as well as 4 equilibrium points near comet 67P/CG. Jiang et al. [31] discussed motions near equilibrium points for a general rotating asteroid.

Investigation of stability of motion in the vicinity of bodies needs to consider the Jacobian integral and zero-velocity surfaces, which can divide the region into the forbidden region and the allowable region for the particle [25,27,30,33]. Besides, stability of equilibrium points determines the motion stabilization near equilibrium points [5,31]; while asymptotically stable manifold, the asymptotically unstable manifold, and the central manifold determine the stability of orbits[31,33]; orbits can be local or large, local orbits [31] are in the vicinity of equilibrium points while large orbits [25,27] are around the whole irregular body.

Periodic orbits in the potential of a rotating body exist and sometimes look sophisticated [2,16,25,30,32-33], which are found that can be classified into different periodic orbit families [25,31,33]. Broucke and Elipe [10] presented 10 families of periodic orbits as examples near a massive ring, and the classification method is the different levels of symmetry with directly geometrical looks. Riaguas et al. [2] found several families of periodic orbits near a massive straight segment with different stability. Jiang et al. [31] classified periodic orbits near equilibrium points into several families on the basis of topological characteristics.

Bifurcations of motion emerge with parameters variation. Riaguas et al. [2] discovered bifurcations when periodic orbits were considered with parameters variation in the gravity field of a massive straight segment. Galán et al. [39] analyzed bifurcations of the figure-8 solution of the three-body problem. Elipe and Lara [5] discussed bifurcations at the 1:1 resonance in the gravity field of a massive straight segment, and found motion near resonance with parameters variation can leads to bifurcations. Resonances have several types: the integer ratio of the orbital angular velocity to the rotational angular velocity of the asteroid can leads to resonances [25,33], in which resonances caused by the 1:1 ratio of rotational period means equilibrium points [22,27]; moreover, the integer ratio of the orbital period to the rotational period of the asteroid can also leads to resonances [28]. Jiang et al. [36] discovered four kinds of bifurcations for the periodic orbits in the potential field of highly irregular-shaped celestial bodies, including tangent bifurcations, period-doubling bifurcations, Neimark-Sacker bifurcations, and the real saddle bifurcations.

Another important dynamical behaviour is the chaos [5,7,18]. Chaos in the rotation and revolution of a massive line segment and a massive point has been discovered by Lindner et al. [7], which include stable synchronous orbit, generic chaotic orbits, sequences of unstable periodic orbits and spin-stabilized orbits. Poincaré surface of sections can also be calculated to analyze the chaos behaviours in the potential of a body [8,10,16,18]. Broucke and Elipe [10] calculated the Poincaré surface of sections in the potential of a solid circular ring, and found out several visible islands, corresponding to tori which surround stable periodic orbits. Blesa [16] ploted Poincaré surface of sections in the potential of a triangular and square plate, respectively. Najid et al. [8] calculated the Poincaré surface of sections to give an overview of the structure of the dynamical behaviour of the test particle in the potential of a massive inhomogeneous straight segment. Besides, Liu et al. [18] used Poincaré surface of sections to find periodic orbits around a homogeneous cube.

The intent of this paper is to provide a detailed look at the order and chaos of the motion near equilibrium points in the potential of a rotating highly irregular-shaped celestial body, with arbitrary shape. In other words, the conclusion is correct for all kinds of simple-shaped or highly irregular-shaped bodies. Sect. 2 presents the motion equation in the potential of a rotating body and the linearized equation of motion relative to the equilibrium point. Sect. 3 studies the order of the motion in the neighborhood of the equilibrium point acting parameter variation. A theorem is presented and proved to describe the existence and continuous of the non-degenerate equilibrium point in the presence of persistently acting parameter variation. The structural stability of non-degenerate and non-resonant equilibrium points is also discussed. Sect. 4 analyzed the topological structure in the vicinity of equilibrium points. The topological structure in the vicinity of equilibrium points has 6 ordinary cases, 3 resonant cases, 3 degenerate real saddle cases, 1 degenerate-equilibrium and resonant case, as well as 1 degenerate-equilibrium and degenerate real saddle case. Besides, structure of submanifolds, stability, number of periodic orbit families as well as the phase diagram of motion near equilibrium points are presented in Sect. 4. Sect. 5 gives a minimum estimate for the number of periodic orbits on a fixed energy hypersurface. Sect. 6 discussed the chaos of the motion in the neighborhood of the equilibrium point acting parameter variation. These conclusions can be applied to the preliminary analysis of the dynamical behaviors around the rotating celestial bodies, spherical or non-spherical shaped, including asteroids, comets, planets and satellites of planets.

Asteroids 216 Kleopatra, 2063 Bacchus, and 25143 Itokawa are selected for applications because the highly irregular-shape of these three bodies. Equilibrium points of asteroids 216 Kleopatra, 2063 Bacchus and 25143 Itokawa exist and move continuously. Eigenvalues affiliated to the equilibrium points for the asteroid 216 Kleopatra move and always belong to the same topological cases when parameter of the rotational angular velocity increasing between $[0.5\omega, 2.0\omega]$. The YORP effect make asteroids' rotation period vary, Asteroid 54509 YORP (2000 PH5)'s rotation speed increased in the past 600,000 years [39]. If the rotation period becomes double or a half of the initial value, what will happen to the equilibrium points? From the parameters of equilibrium points with the rotation speed's interval $[0.5\omega, 2.0\omega]$, one can see the changes of location and eigenvalues of equilibrium points. Besides, when the parameter varies out of the interval, locations of equilibrium points for 216 Kleopatra enter the body of the asteroid. So we choose the interval $[0.5\omega, 2.0\omega]$. Eigenvalues affiliated to the equilibrium point E3 and E4 for the asteroid 2063 Bacchus and 25143 Itokawa will collide and move though the resonant cases with the same region of rotational angular velocity : $[0.5\omega, 2.0\omega]$. Poincaré sections in the potential of the asteroid 216 Kleopatra are presented to show the chaos behaviors in the large, yields the chaos occurs not only locally, but also largely.

## 2. Equations of Motion

The body-fixed frame is established with the origin located at the body's mass center, the z axis coincides with the rotation axis, the x axis and y axis coincide with the other two principal axes of the body. The dynamical model used across this paper is the following equation [36], which depends on parameters

$$\ddot{\mathbf{r}} + 2\boldsymbol{\omega}(\boldsymbol{\mu}_\omega) \times \dot{\mathbf{r}} + \dot{\boldsymbol{\omega}}(\boldsymbol{\mu}_\omega) \times \mathbf{r} + \frac{\partial V(\boldsymbol{\mu}, \mathbf{r})}{\partial \mathbf{r}} = 0, \qquad (1)$$

where $\mathbf{r}$ is the body-fixed vector from the body center of mass to the particle, $\boldsymbol{\omega}$ is the rotation angular velocity vector of the body relative to the inertial space, $U(\mathbf{r})$ is the gravitational potential, $V(\boldsymbol{\mu}, \mathbf{r}) = -\frac{1}{2}(\boldsymbol{\omega}(\boldsymbol{\mu}_\omega) \times \mathbf{r})(\boldsymbol{\omega}(\boldsymbol{\mu}_\omega) \times \mathbf{r}) + U(\boldsymbol{\mu}_U, \mathbf{r})$, $\boldsymbol{\mu}_\omega = \boldsymbol{\mu}_\omega(t)$ is the parameter for $\boldsymbol{\omega}$, $\boldsymbol{\mu}_U = \boldsymbol{\mu}_U(t)$ is the parameter for $U(\mathbf{r})$, $\boldsymbol{\mu} = \boldsymbol{\mu}(t) = [\boldsymbol{\mu}_\omega(t), \boldsymbol{\mu}_U(t)]$, $\boldsymbol{\mu}$ is the parameter related to the time $t$. The dimension of $\boldsymbol{\mu}_\omega = \boldsymbol{\mu}_\omega(t)$ or $\boldsymbol{\mu}_U = \boldsymbol{\mu}_U(t)$ is arbitrary, it can be 0, 1, 2, etc. The physical meaning of the parameter $\boldsymbol{\mu}_\omega$ represents the change of the rotation rate of the body, while the physical meaning of the parameter $\boldsymbol{\mu}_U$ represents the change of the shape and structure of the body. There are several phenomena that can make the parameter $\boldsymbol{\mu}$ varies, including the YORP effect [40], the generating of the dust tails from the cometary nucleus [41], the surface grain motion [42], the disintegration of asteroids [43], etc.

Denote $\mathbf{F}(\boldsymbol{\mu},\mathbf{r}) = \dfrac{\partial V(\boldsymbol{\mu},\mathbf{r})}{\partial \mathbf{r}}$.

The Jacobian integral [33,44] is

$$\frac{1}{2}\dot{\mathbf{r}}\cdot\dot{\mathbf{r}} + V(\mathbf{r}) = H, \tag{2}$$

The equilibrium points satisfy the following condition

$$\frac{\partial V(\boldsymbol{\mu},\mathbf{r})}{\partial x} = \frac{\partial V(\boldsymbol{\mu},\mathbf{r})}{\partial y} = \frac{\partial V(\boldsymbol{\mu},\mathbf{r})}{\partial z} = 0 \tag{3}$$

where $(x,y,z)$ is the component of $\mathbf{r}$ in the body-fixed coordinate system. Let $(x_L, y_L, z_L)^T$ denote the coordinate of an equilibrium point, the effective potential $V(x,y,z)$ can be expanded by Taylor series at the equilibrium point $(x_L, y_L, z_L)^T$.

The characteristic equation [31] of the equilibrium point is

$$\lambda^6 + \left(V_{xx}+V_{yy}+V_{zz}+4\omega^2\right)\lambda^4 + \left(V_{xx}V_{yy}+V_{yy}V_{zz}+V_{zz}V_{xx}-V_{xy}^2-V_{yz}^2-V_{xz}^2+4\omega^2 V_{zz}\right)\lambda^2 \\ + \left(V_{xx}V_{yy}V_{zz}+2V_{xy}V_{yz}V_{xz}-V_{xx}V_{yz}^2-V_{yy}V_{xz}^2-V_{zz}V_{xy}^2\right)=0 \tag{4}$$

where $\lambda$ denote the eigenvalues of the equilibrium point. Denote the Jacobi constant at the equilibrium point as $H(L)$, and the eigenvector of the eigenvalue $\lambda_j$ as $\mathbf{u}_j$.

## 3. Order in the Neighborhood of the Equilibrium Point with Variable Parameters

In this section, we discuss the order of the motion in the neighborhood of the equilibrium point around a rotating body with variable parameters, including the existence and continuous of equilibrium points acting parameter variation as well as the structural stability of non-degenerate and non-resonant equilibrium points.

**3.1 Existence and continuous of equilibrium points when parameter changes**

The Hessian matrix of the effective potential at the equilibrium point $\mathbf{r} = \boldsymbol{\tau}(\boldsymbol{\mu}_0) = \boldsymbol{\tau}_0$ is $\nabla^2 V(\boldsymbol{\mu}_0, \boldsymbol{\tau}_0) = \dfrac{\partial \mathbf{F}(\boldsymbol{\mu}, \boldsymbol{\tau})}{\partial \boldsymbol{\tau}}\bigg|_{(\boldsymbol{\mu}_0, \boldsymbol{\tau}_0)}$. The non-degenerate equilibrium point is the equilibrium point with the Hessian matrix of the effective potential that has full rank.

**Definition 1.** Assume that when $\boldsymbol{\mu} = \boldsymbol{\mu}_0$, the point $\mathbf{r} = \boldsymbol{\tau}(\boldsymbol{\mu}_0) = \boldsymbol{\tau}_0$ is an equilibrium point of the motion for the particle in the potential field of a rotating body. If there exists an open neighborhood of $\boldsymbol{\mu}_0$, which can be denoted as $G_N(\boldsymbol{\mu}_0)$, where $N$ in $G_N(\boldsymbol{\mu}_0)$ means neighborhood; and there exists a function $\boldsymbol{\tau} = \boldsymbol{\tau}(\boldsymbol{\mu})$, which satisfies $\boldsymbol{\tau}(\boldsymbol{\mu}_0) = \boldsymbol{\tau}_0$; such that $\forall \boldsymbol{\mu} \in G_N(\boldsymbol{\mu}_0)$, $\boldsymbol{\tau} = \boldsymbol{\tau}(\boldsymbol{\mu})$ is an equilibrium point. Then the equilibrium point $\mathbf{r} = \boldsymbol{\tau}(\boldsymbol{\mu}_0) = \boldsymbol{\tau}_0$ is said to be *continuous in the presence of persistently acting parameter variation.*

**Theorem 1.** The non-degenerate equilibrium point is continuous in the presence of persistently acting parameter variation.

**Proof.** The equilibrium point $\boldsymbol{\tau} = \boldsymbol{\tau}(\boldsymbol{\mu})$ is the implicit function expressed by $\mathbf{F}(\boldsymbol{\mu}, \boldsymbol{\tau}) = 0$. Consider a non-degenerate equilibrium point $\mathbf{r} = \boldsymbol{\tau}(\boldsymbol{\mu}_0) = \boldsymbol{\tau}_0$, it satisfies $\mathbf{F}(\boldsymbol{\mu}_0, \boldsymbol{\tau}_0) = 0$, and the matrix $\nabla^2 V(\boldsymbol{\mu}_0, \boldsymbol{\tau}_0) = \dfrac{\partial \mathbf{F}(\boldsymbol{\mu}, \boldsymbol{\tau})}{\partial \boldsymbol{\tau}}\bigg|_{(\boldsymbol{\mu}_0, \boldsymbol{\tau}_0)}$ has full rank. Apply the implicit function theorem, there is a continuous function $\mathbf{r} = \boldsymbol{\tau}(\boldsymbol{\mu})$ such that $\boldsymbol{\tau}(\boldsymbol{\mu}_0) = \boldsymbol{\tau}_0$, and $\mathbf{F}(\boldsymbol{\mu}, \boldsymbol{\tau}(\boldsymbol{\mu})) = 0$ for any $\boldsymbol{\mu} \in G_N(\boldsymbol{\mu}_0)$, where $G_N(\boldsymbol{\mu}_0)$ is a sufficiently small open neighborhood of $\boldsymbol{\mu}_0$. □

From the definition of the effective potential, one can see that the effective potential is a continuous function about the parameter $\boldsymbol{\mu}$, that means if the parameter $\boldsymbol{\mu}$ has a sufficiently small change, the effective potential has a sufficiently small change. From Theorem 1, it is shown that if the equilibrium point is non-degenerate, the equilibrium point exists with a sufficiently small deviation of the effective potential. Besides, the effective potential $V$ at the position $\mathbf{r}$ relative to the body-fixed frame relates to the gravitational potential $U$ and the rotational angular velocity vector of the body relative to the inertial space $\boldsymbol{\omega}$. In other words, if the equilibrium point is non-degenerate, the equilibrium point exists with a sufficiently small deviation of the gravitational field of the body and the rotation angular velocity. This leads to Remark 1 and Remark 2.

**Remark 1.** If the equilibrium point is non-degenerate, the equilibrium point exists with a sufficiently small deviation of the effective potential.

**Remark 2.** If the equilibrium point is non-degenerate, the equilibrium point exists with a sufficiently small deviation of the gravitational field of the body and the rotation angular velocity.

Using the Theorem 1, one can know that if the parameter $\boldsymbol{\mu}$ is continuously changing in a sufficiently small open neighborhood, the implicit function which is defined by the equation $\mathbf{F}(\boldsymbol{\mu}, \boldsymbol{\tau}(\boldsymbol{\mu})) = 0$ is a continuous function, then the equilibrium point $\boldsymbol{\tau} = \boldsymbol{\tau}(\boldsymbol{\mu})$ is continuously changing. This leads to Remark 3.

**Remark 3.** If the parameter $\boldsymbol{\mu}$ is continuously changing in a sufficiently small open neighborhood, then the equilibrium point $\boldsymbol{\tau} = \boldsymbol{\tau}(\boldsymbol{\mu})$ is continuously changing.

## 3.2 Structural stability of Non-Degenerate and Non-Resonant Equilibrium Points

In this section, we discuss the stability of the equilibrium point acting parameter variation. If a small body or a spacecraft is placed at the equilibrium point or near the equilibrium point, the motion of the small body or the spacecraft is relating to the existence, continuous, and stability of the equilibrium point.

The non-resonant equilibrium point is the equilibrium point with the eigenvalues that are linearly independent, while the resonant equilibrium point is corresponding to the eigenvalues that are linearly dependent. The non-degenerate and non-resonant equilibrium points include linear stable equilibrium points and non-resonant unstable equilibrium points.

**Theorem 2.** If the non-degenerate equilibrium point $\mathbf{r}_0 = \boldsymbol{\tau}(\boldsymbol{\mu}_0) = \boldsymbol{\tau}_0$ in the gravitational field of a rotating body is linear stable, then there is a sufficiently small open neighborhood of $\boldsymbol{\mu}_0$, which can be denoted as $G_N(\boldsymbol{\mu}_0)$, such that for any $\boldsymbol{\mu} \in G_N(\boldsymbol{\mu}_0)$, the non-degenerate equilibrium point exists and is linear stable. In other words, there is a function $\mathbf{r} = \boldsymbol{\tau}(\boldsymbol{\mu})$ such that $\boldsymbol{\tau}(\boldsymbol{\mu}_0) = \boldsymbol{\tau}_0$, $\mathbf{F}(\boldsymbol{\mu},\mathbf{r}) = \dfrac{\partial V(\boldsymbol{\mu},\mathbf{r})}{\partial \mathbf{r}} = 0$ is established for any $\boldsymbol{\mu} \in G_N(\boldsymbol{\mu}_0)$, and $\mathbf{r} = \boldsymbol{\tau}(\boldsymbol{\mu})$ is the non-degenerate equilibrium point of Eq. (1), in addition, the equilibrium point $\mathbf{r} = \boldsymbol{\tau}(\boldsymbol{\mu})$ is linear stable.

Furthermore, if the non-degenerate equilibrium point is non-resonant, the conclusion about the structural stability is similar with Theorem 2, which is

**Theorem 3.** If the non-degenerate equilibrium point $\mathbf{r}_0 = \boldsymbol{\tau}(\boldsymbol{\mu}_0) = \boldsymbol{\tau}_0$ in the gravitational field of a rotating body is non-resonant, then there is a sufficiently small open neighborhood of $\boldsymbol{\mu}_0$, which can be denoted as $G_N(\boldsymbol{\mu}_0)$, such that for any $\mu \in G_N(\mu_0)$, the non-degenerate equilibrium point exists and the stability of the equilibrium point is invariant. That means the manifolds near the equilibrium point $\mathbf{r} = \tau(\mu)$ is diffeomorphic with the manifolds near the equilibrium point $\mathbf{r}_0 = \tau(\mu_0)$. In other words, there is a function $\mathbf{r} = \tau(\mu)$ such that $\tau(\mu_0) = \tau_0$, $\mathbf{F}(\mu, \mathbf{r}) = \dfrac{\partial V(\mu, \mathbf{r})}{\partial \mathbf{r}} = 0$ is established for any $\mu \in G_N(\mu_0)$, and $\mathbf{r} = \tau(\mu)$ is the non-degenerate equilibrium point of Eq. (1), in addition, the manifolds near the equilibrium point $\mathbf{r} = \tau(\mu)$ is diffeomorphic with the manifolds near the equilibrium point $\mathbf{r}_0 = \tau(\mu_0)$.

Considering Theorem 2 can be obtained from Theorem 3, the proof for Theorem 2 is omitted, and the proof for Theorem 3 is presented as follows:

**Proof for Theorem 3.** The non-degenerate and non-resonant equilibrium point $\tau = \tau(\mu)$ is the implicit function expressed by $\mathbf{F}(\mu, \tau) = 0$. The equilibrium point satisfies $\mathbf{F}(\mu_0, \tau_0) = 0$, besides the matrix $\nabla^2 V(\mu_0, \tau_0) = \dfrac{\partial \mathbf{F}(\mu, \tau)}{\partial \tau}\bigg|_{(\mu_0, \tau_0)}$ has full rank. The eigenvalues can be expressed as the implicit function defined by the characteristic equation $P(\lambda) = 0$, where $\lambda$ is an eigenvalue. Apply the implicit function theorem to the implicit function defined by the characteristic equation $P(\lambda) = 0$. If one eigenvalue $\lambda_i(\mu_0)$ has non-zero real part, then for any $\mu \in G_N(\mu_0)$, where $G_N(\mu_0)$ is a sufficiently small open neighborhood of $\mu_0$, the eigenvalue $\lambda_i(\mu)$ also has non-zero real part. The sufficiently small open neighborhood $G_N(\mu_0)$ exists because the determinant $|\nabla^2 V(\mu_0, \tau_0)| \neq 0$, while $|\nabla^2 V(\mu_0, \tau_0)| \neq 0$ because the matrix $\nabla^2 V(\mu_0, \tau_0) = \dfrac{\partial \mathbf{F}(\mu, \tau)}{\partial \tau}\bigg|_{(\mu_0, \tau_0)}$ has full rank.

Moreover, for any $\mathbf{\mu} \in G_N(\mathbf{\mu}_0)$, where $G_N(\mathbf{\mu}_0)$ is a sufficiently small open neighborhood of $\mathbf{\mu}_0$, the pure imaginary eigenvalue $\lambda_i(\mathbf{\mu}_0)$ becomes to $\lambda_i(\mathbf{\mu})$, and $\lambda_i(\mathbf{\mu})$ is also a pure imaginary eigenvalue. Using the inverse function theorem [45], we have conclude that the manifolds near the equilibrium point $\mathbf{r} = \mathbf{\tau}(\mathbf{\mu})$ is diffeomorphic with the manifolds near the equilibrium point $\mathbf{r}_0 = \mathbf{\tau}(\mathbf{\mu}_0)$. $\square$

### 3.3 Number of Periodic Orbits on a fixed Energy Hypersurface

On the manifold $H = h \left( h = H(L) + \varepsilon^2 \right)$, which is a fixed 5-dimentional energy hypersurface, how many periodic orbits exist at least?

Denote the Hessian matrix of the Jacobian integral at the equilibrium point L as

$$H_{\mathbf{rr}} = \begin{pmatrix} H_{xx} & H_{xy} & H_{xz} \\ H_{xy} & H_{yy} & H_{yz} \\ H_{xz} & H_{yz} & H_{zz} \end{pmatrix}, \tag{5}$$

Where $H_{uv} = \left( \dfrac{\partial^2 H}{\partial u \partial v} \right)_L$, in which $(u, v = x, y, z)$.

From Eq. (3), one can obtain that the Hessian matrix of the Jacobian integral equals to the Hessian matrix of the effective potential, then:

$$H_{\mathbf{rr}} = \begin{pmatrix} H_{xx} & H_{xy} & H_{xz} \\ H_{xy} & H_{yy} & H_{yz} \\ H_{xz} & H_{yz} & H_{zz} \end{pmatrix} = V_{\mathbf{rr}} \triangleq \begin{pmatrix} V_{xx} & V_{xy} & V_{xz} \\ V_{xy} & V_{yy} & V_{yz} \\ V_{xz} & V_{yz} & V_{zz} \end{pmatrix} \tag{6}$$

If the Hessian matrix of the Jacobian integral for an equilibrium point in the potential field of a rotating body is positive definite, there are at least three periodic orbits on the energy hypersurface $h = H(L) + \varepsilon^2$ [31,46-47]. However, the positive definite condition [31] for the Hessian matrix of the Jacobian integral or effective potential is only a sufficient condition for the stability of the equilibrium points in the potential field of a rotating body, not a sufficient and necessary condition. Consider the Hessian matrix $H_{rr} = V_{rr}$ restricted to $E^c(L)$ (where $T_L\mathbf{S} \cong E^e(L) \oplus E^s(L) \oplus E^c(L) \oplus E^u(L)$ and $\dim E^c(L) \neq 0$), which can be not positive definite, there are at least $\frac{1}{2}\left[\dim E^c(L) - \dim E^r(L)\right] + \text{sgn}\left[\dim E^r(L)\right]$ families of periodic orbits near the equilibrium point, where $\text{sgn}(\alpha) = \begin{cases} 1, & \alpha > 0 \\ 0, & \alpha = 0 \end{cases}$; this leads to at least $\frac{1}{2}\left[\dim E^c(L) - \dim E^r(L)\right] + \text{sgn}\left[\dim E^r(L)\right]$ periodic orbits on the energy hypersurface $h = H(L) + \varepsilon^2$.

## 4. Topological Structure in the Vicinity of Equilibrium Points

Denote $M$ as the topological space generated by $(x, y, z)$ which has a metric. Let the tangent space of the equilibrium point $L \in M$ be $T_L M$, and the sufficiently small open neighbourhood of the equilibrium point on the smooth manifold $M$ be $\Xi$. Define the tangent bundle as

$$TM = \bigcup_{p \in M} T_p M = \{(p, q) | p \in M, q \in T_p M\}$$

and

$$T\Xi = \bigcup_{p \in \Xi} T_p \Xi = \{(p, q) | p \in \Xi, q \in T_p \Xi\},$$

which satisfy $\dim TM = \dim T\Xi = 6$; Let $(\mathbf{S}, \Omega)$ be the 6-dimensional symplectic manifold near the equilibrium point, where $\Omega$ is a non-degenerate skew-symmetric bilinear quadratic form.

Denote $C_\lambda = \{\lambda \in \mathbb{C} | \lambda \text{ is the root of Eq. (4)}\}$. Let the Jacobian constant at the equilibrium point be $H(L)$, the eigenvector of the eigenvalue $\lambda_j$ be $\mathbf{u}_j$.

Denote

$$E^s(L) = span\{\mathbf{u}_j | \operatorname{Re}\lambda_j < 0\},$$

$$E^c(L) = span\{\mathbf{u}_j | \operatorname{Re}\lambda_j = 0\},$$

$$E^u(L) = span\{\mathbf{u}_j | \operatorname{Re}\lambda_j > 0\}.$$

$$E^e(L) = span\{\mathbf{u}_j | \lambda_j = 0\}$$

On the manifold $H = h\left(h = H(L) + \varepsilon^2\right)$, the asymptotically stable manifold $W^s(\mathbf{S})$, the asymptotically unstable manifold $W^u(\mathbf{S})$, the central manifold $W^c(\mathbf{S})$ and the degenerate equilibrium manifold $W^e(\mathbf{S})$ near the equilibrium point are tangent to the asymptotically stable subspace $E^s(L)$, the asymptotically unstable subspace $E^u(L)$, the central subspace $E^c(L)$ and the degenerate equilibrium subspace $E^e(L)$ at the equilibrium point, respectively; where $\varepsilon^2$ is sufficiently small that there is no other equilibrium point $\tilde{L}$ in the sufficiently small open neighbourhood on the manifold $(\mathbf{S}, \Omega)$ with the Jacobian constant $H(\tilde{L})$ that satisfies $H(L) \leq H(\tilde{L}) \leq h$.

Denote 
$$\begin{cases} \bar{E}^s(L) = span\{\mathbf{u}_j | \lambda_j \in C_\lambda, \operatorname{Re}\lambda_j < 0, \operatorname{Im}\lambda_j = 0\} \\ \tilde{E}^s(L) = span\{\mathbf{u}_j | \lambda_j \in C_\lambda, \operatorname{Re}\lambda_j < 0, \operatorname{Im}\lambda_j \neq 0\} \end{cases}$$ and

$$\begin{cases} \bar{E}^u(L) = span\{\mathbf{u}_j | \lambda_j \in C_\lambda, \operatorname{Re}\lambda_j > 0, \operatorname{Im}\lambda_j = 0\} \\ \tilde{E}^u(L) = span\{\mathbf{u}_j | \lambda_j \in C_\lambda, \operatorname{Re}\lambda_j > 0, \operatorname{Im}\lambda_j \neq 0\} \end{cases}.$$

The asymptotically stable manifold $\bar{W}^s(\mathbf{S})$ and $\tilde{W}^s(\mathbf{S})$ are tangent to the asymptotically stable subspace $\bar{E}^s(L)$ and $\tilde{E}^s(L)$ at the equilibrium point, respectively. The asymptotically unstable manifold $\bar{W}^u(\mathbf{S})$ and $\tilde{W}^u(\mathbf{S})$ are tangent to the asymptotically unstable subspace $\bar{E}^u(L)$ and $\tilde{E}^u(L)$ at the equilibrium point, respectively.

Denote $W^r(\mathbf{S})$ as the resonant manifold, which is tangent to the resonant subspace $E^r(L) = span\{\mathbf{u}_j | \exists \lambda_k, s.t. \operatorname{Re} \lambda_j = \operatorname{Re} \lambda_k = 0, \operatorname{Im} \lambda_j = \operatorname{Im} \lambda_k, j \neq k\}$, it is clear that $E^r(L) \subseteq E^c(L)$ and $W^r(\mathbf{S}) \subseteq W^c(\mathbf{S})$. Denote $W^f(\mathbf{S})$ as the uniform manifold, which is tangent to the uniform subspace $E^f(L) = span\{\mathbf{u}_j | \lambda_j \in C_\lambda, \exists \lambda_k, s.t. \operatorname{Re} \lambda_j = \operatorname{Re} \lambda_k \neq 0, \operatorname{Im} \lambda_j = \operatorname{Im} \lambda_k = 0, j \neq k\}$. When $\dim E^f(L) \neq 0$, the manifolds and subspaces of $\lambda_j$ and $\lambda_k$ are uniform and their phase diagrams are coincident.

Denote $E^l(L) = span\{\mathbf{u}_j | \operatorname{Re} \lambda_j = 0, \operatorname{Im} \lambda_j \neq 0, s.t. \forall \lambda_k \in C_\lambda, \lambda_k \neq \lambda_j\}$ as the linear stable space, where $J$ is the number of $\lambda_j$ which satisfies $\begin{cases} \operatorname{Re} \lambda_j = 0 \\ \operatorname{Im} \lambda_j \neq 0 \end{cases}$, $Z$ is the set of integer. Denote $W^l(\mathbf{S})$ as the linear stable manifold, which is tangent to the linear stable subspace, it yields $E^l(L) \subseteq E^c(L)$ and $W^l(\mathbf{S}) \subseteq W^c(\mathbf{S})$.

Denote $\simeq$ as the topological homeomorphism, $\cong$ as the diffeomorphism and $\oplus$ as the direct sum. Denote $T_L \mathbf{S}$ as the tangent space of the manifold $(\mathbf{S}, \Omega)$. Then, $(\mathbf{S}, \Omega) \simeq T\Xi \cong W^s(\mathbf{S}) \oplus W^c(\mathbf{S}) \oplus W^u(\mathbf{S})$ and $T_L \mathbf{S} \cong E^s(L) \oplus E^c(L) \oplus E^u(L)$, the topological structure of the submanifolds fixes the phase diagram of the motion around equilibrium points [31].

Based on the conclusion above, one can get the following theorem. It is about the topological classification of the equilibrium points in the potential field of a rotating body.

**Theorem 4.** Consider the topological structure of the manifolds in the vicinity of the equilibrium points in the potential field of a rotating body. There exist 6 ordinary cases, 3 resonant cases, 3 degenerate real saddle cases, 7 degenerate-equilibrium cases, 1 degenerate-equilibrium and resonant case, as well as 1 degenerate-equilibrium and degenerate real saddle case. Classifications and properties of these cases are shown as follows and in Figure 1:

**I. Ordinary Cases:**

**Case O1:** The eigenvalues are in the form of

$\pm i\beta_j \left(\beta_j \in \mathrm{R}, \beta_j > 0; j=1,2,3 | \forall k \neq j, k=1,2,3, s.t. \beta_k \neq \beta_j \right)$; then, the structure of the submanifold is $(\mathbf{S},\Omega) \simeq T\Xi \cong W^c(\mathbf{S})$, and $\dim W^r(\mathbf{S}) = 0$.

**Case O2:** The forms of the eigenvalues are $\pm \alpha_j \left(\alpha_j \in \mathrm{R}, \alpha_j > 0, j=1\right)$ and $\pm i\beta_j \left(\beta_j \in \mathrm{R}, \beta_j > 0; j=1,2 | \beta_1 \neq \beta_2 \right)$; then, the structure of the submanifold is $(\mathbf{S},\Omega) \simeq T\Xi \cong \overline{W}^s(\mathbf{S}) \oplus W^c(\mathbf{S}) \oplus \overline{W}^u(\mathbf{S})$, $\dim W^c(\mathbf{S}) = 4$, $\dim \overline{W}^s(\mathbf{S}) = \dim \overline{W}^u(\mathbf{S}) = 1$, and $\dim W^r(\mathbf{S}) = 0$.

**Case O3:** The forms of the eigenvalues are $\pm \alpha_j \left(\alpha_j \in \mathrm{R}, \alpha_j > 0; j=1,2 | \alpha_1 \neq \alpha_2\right)$ and $\pm i\beta_j \left(\beta_j \in \mathrm{R}, \beta_j > 0, j=1 \right)$; then, the structure of the submanifold is

$(\mathbf{S},\Omega) \simeq T\Xi \cong \overline{W}^s(\mathbf{S}) \oplus W^c(\mathbf{S}) \oplus \overline{W}^u(\mathbf{S})$, $\dim W^r(\mathbf{S}) = 0$, and

$\dim \overline{W}^s(\mathbf{S}) = \dim W^c(\mathbf{S}) = \dim \overline{W}^u(\mathbf{S}) = 2$.

**Case O4:** The forms of the eigenvalues are $\pm i\beta_j \left(\beta_j \in \mathrm{R}, \beta_j > 0, j=1\right)$ and $\pm \sigma \pm i\tau \left(\sigma, \tau \in \mathrm{R}; \sigma, \tau > 0 \right)$; then, the structure of the submanifold is

$(\mathbf{S},\Omega) \simeq T\Xi \cong \tilde{W}^s(\mathbf{S}) \oplus W^c(\mathbf{S}) \oplus \tilde{W}^u(\mathbf{S})$, $\dim W^r(\mathbf{S}) = 0$, and

$\dim \tilde{W}^s(\mathbf{S}) = \dim W^c(\mathbf{S}) = \dim \tilde{W}^u(\mathbf{S}) = 2$.

**Case O5:** The forms of the eigenvalues are $\pm\alpha_j\,(\alpha_j \in \mathrm{R}, \alpha_j > 0, j=1)$ and $\pm\sigma \pm i\tau\,(\sigma,\tau \in \mathrm{R}; \sigma,\tau > 0)$; then, the structure of the submanifold is $(\mathbf{S},\Omega) \simeq T\Xi \cong \bar{W}^s(\mathbf{S}) \oplus \bar{W}^u(\mathbf{S}) \oplus \tilde{W}^s(\mathbf{S}) \oplus \tilde{W}^u(\mathbf{S})$, and $\dim \bar{W}^s(\mathbf{S}) = \dim \bar{W}^u(\mathbf{S}) = 1$ $\dim \tilde{W}^s(\mathbf{S}) = \dim \tilde{W}^u(\mathbf{S}) = 2$.

**Case O6:** The forms of the eigenvalues are

$\pm\alpha_j\,(\alpha_j \in \mathrm{R}, \alpha_j > 0, j=1,2,3 | \forall k \neq j, k=1,2,3, s.t. \alpha_k \neq \alpha_j)$; then, the structure of the submanifold is $(\mathbf{S},\Omega) \simeq T\Xi \cong \bar{W}^s(\mathbf{S}) \oplus \bar{W}^u(\mathbf{S})$, and $\dim \bar{W}^s(\mathbf{S}) = \dim \bar{W}^u(\mathbf{S}) = 3$.

**II. Resonant Cases:**

**Case R1:** The forms of the eigenvalues are $\pm i\beta_j\,(\beta_j \in \mathrm{R}, \beta_1 = \beta_2 = \beta_3 > 0; j=1,2,3)$; then, the structure of the submanifold is $(\mathbf{S},\Omega) \simeq T\Xi \simeq W^c(\mathbf{S}) \simeq W^r(\mathbf{S})$, and $\dim W^r(\mathbf{S}) = \dim W^c(\mathbf{S}) = 6$.

**Case R2:** The forms of the eigenvalues are

$\pm i\beta_j\,(\beta_j \in \mathrm{R}, \beta_j > 0, \beta_1 = \beta_2 \neq \beta_3; j=1,2,3)$; then, the structure of the submanifold is $(\mathbf{S},\Omega) \simeq T\Xi \cong W^c(\mathbf{S})$, and $\dim W^r(\mathbf{S}) = 4$.

**Case R3:** The forms of the eigenvalues are

$\pm\alpha_j\,(\alpha_j \in \mathrm{R}, \alpha_j > 0, j=1), \pm i\beta_j\,(\beta_j \in \mathrm{R}, \beta_1 = \beta_2 > 0; j=1,2)$; then, the structure of the submanifold is $(\mathbf{S},\Omega) \simeq T\Xi \cong \bar{W}^s(\mathbf{S}) \oplus W^c(\mathbf{S}) \oplus \bar{W}^u(\mathbf{S})$, $\dim W^r(\mathbf{S}) = \dim W^c(\mathbf{S}) = 4$, and $\dim \bar{W}^s(\mathbf{S}) = \dim \bar{W}^u(\mathbf{S}) = 1$.

**III. Degenerate Real Saddle Cases**

**Case DRS1:** The forms of the eigenvalues are

$\pm\alpha_j\,(\alpha_j \in \mathrm{R}, \alpha_j > 0, j=1,2,3 | \alpha_1 = \alpha_2 \neq \alpha_3)$; then, the structure of the submanifold is

$(S, \Omega) \simeq T\Xi \cong \bar{W}^s(S) \oplus \bar{W}^u(S)$, and $\dim \bar{W}^s(S) = \dim \bar{W}^u(S) = 3$.

**Case DRS2:** The forms of the eigenvalues are

$\pm \alpha_j \left( \alpha_j \in R, \alpha_j > 0, j = 1,2,3 | \alpha_1 = \alpha_2 = \alpha_3 \right)$; then, the structure of the submanifold is

$(S, \Omega) \simeq T\Xi \cong \bar{W}^s(S) \oplus \bar{W}^u(S)$, and $\dim \bar{W}^s(S) = \dim \bar{W}^u(S) = 3$.

**Case DRS3:** The forms of the eigenvalues are

$\pm i\beta_j \left( \beta_j \in R, \beta_j > 0, j = 1 \right)$ and $\pm \alpha_j \left( \alpha_j \in R, \alpha_j > 0, j = 1,2 | \alpha_1 = \alpha_2 \right)$; then, the

structure of the submanifold is $(S, \Omega) \simeq T\Xi \cong \bar{W}^s(S) \oplus W^c(S) \oplus \bar{W}^u(S)$, and

$\dim \bar{W}^s(S) = \dim \bar{W}^u(S) = 2$.

**IV. Degenerate Equilibrium Cases**

**Case DE1:** The forms of the eigenvalues are $\gamma_j \left( \gamma_j = 0; j = 1,2 \right)$ and

$\pm \sigma \pm i\tau \left( \sigma, \tau \in R; \sigma, \tau > 0 \right)$; then, the structure of the submanifold is

$(S, \Omega) \simeq T\Xi \cong W^e(S) \oplus \tilde{W}^s(S) \oplus \tilde{W}^u(S)$, and

$\dim W^e(S) = \dim \tilde{W}^s(S) = \dim \tilde{W}^u(S) = 2$.

**Case DE2:** The forms of the eigenvalues are $\gamma_j \left( \gamma_j = 0; j = 1,2 \right)$ and

$\pm i\beta_j \left( \beta_j \in R, \beta_j > 0; j = 1,2 | \beta_1 \neq \beta_2 \right)$; then, the structure of the submanifold

is $(S, \Omega) \simeq T\Xi \cong W^e(S) \oplus W^c(S)$, $\dim W^e(S) = 2$, $\dim W^c(S) = 4$, and

$\dim W^r(S) = 0$.

**Case DE3:** The forms of the eigenvalues are $\gamma_j \left( \gamma_j = 0; j = 1,2 \right)$ and

$\pm \alpha_j \left( \alpha_j \in R, \alpha_j > 0, j = 1,2 | \alpha_1 \neq \alpha_2 \right)$; then, the structure of the submanifold is

$(S, \Omega) \simeq T\Xi \cong W^e(S) \oplus \bar{W}^s(S) \oplus \bar{W}^u(S)$, $\dim W^s(S) = \dim \bar{W}^s(S) = \dim \bar{W}^u(S) = 2$.

**Case DE4:** The forms of the eigenvalues are $\gamma_j \left( \gamma_j = 0; j = 1,2 \right)$,

$\pm i\beta_1\,(\beta_1 \in \mathrm{R}, \beta_1 > 0)$ and $\pm\alpha_1\,(\alpha_1 \in \mathrm{R}, \alpha_1 > 0)$; then, the structure of the submanifold is $(\mathbf{S},\Omega) \simeq T\Xi \cong W^e(\mathbf{S}) \oplus \overline{W}^s(\mathbf{S}) \oplus W^c(\mathbf{S}) \oplus \overline{W}^u(\mathbf{S})$, $\dim W^e(\mathbf{S}) = \dim W^c(\mathbf{S}) = 2$, $\dim W^s(\mathbf{S}) = \dim W^u(\mathbf{S}) = 1$, and $\dim W^r(\mathbf{S}) = 0$.

**Case DE5:** The forms of the eigenvalues are $\gamma_j\,(\gamma_j = 0;\, j = 1,2,3,4)$ and $\pm i\beta_1\,(\beta_1 \in \mathrm{R}, \beta_1 > 0)$; then, the structure of the submanifold is $(\mathbf{S},\Omega) \simeq T\Xi \cong W^e(\mathbf{S}) \oplus W^c(\mathbf{S})$, $\dim W^c(\mathbf{S}) = 4$, $\dim W^e(\mathbf{S}) = 2$, and $\dim W^r(\mathbf{S}) = 0$.

**Case DE6:** The forms of the eigenvalues are $\gamma_j\,(\gamma_j = 0;\, j = 1,2,3,4)$ and $\pm\alpha_1\,(\alpha_1 \in \mathrm{R}, \alpha_1 > 0)$; then, the structure of the submanifold is $(\mathbf{S},\Omega) \simeq T\Xi \cong W^e(\mathbf{S}) \oplus \overline{W}^s(\mathbf{S}) \oplus \overline{W}^u(\mathbf{S})$, $\dim W^e(\mathbf{S}) = 4$, $\dim \overline{W}^s(\mathbf{S}) = \dim \overline{W}^u(\mathbf{S}) = 1$.

**Case DE7:** The forms of the eigenvalues are $\gamma_j\,(\gamma_j = 0;\, j = 1,2,3,4,5,6)$; then, the structure of the submanifold is $(\mathbf{S},\Omega) \simeq T\Xi \cong W^e(\mathbf{S})$, $\dim W^e(\mathbf{S}) = 6$.

## V. Degenerate Equilibrium and Resonant Case

**Case DER1:** The forms of the eigenvalues are $\gamma_j\,(\gamma_j = 0;\, j = 1,2)$ and $\pm i\beta_j\,(\beta_j \in \mathrm{R}, \beta_1 = \beta_2 > 0;\, j = 1,2)$; then, the structure of the submanifold is $(\mathbf{S},\Omega) \simeq T\Xi \cong W^e(\mathbf{S}) \oplus W^c(\mathbf{S}) \cong W^e(\mathbf{S}) \oplus W^r(\mathbf{S})$, $\dim W^e(\mathbf{S}) = 2$, $\dim W^c(\mathbf{S}) = \dim W^r(\mathbf{S}) = 4$.

## VI. Degenerate Equilibrium and Degenerate Real Saddle Case

**Case DEDRS1:** The forms of the eigenvalues are $\gamma_j\,(\gamma_j = 0;\, j = 1,2)$ and $\pm\alpha_j\,(\alpha_j \in \mathrm{R}, \alpha_j > 0,\, j = 1,2\,|\,\alpha_1 = \alpha_2)$; then, the structure of the submanifold is $(\mathbf{S},\Omega) \simeq T\Xi \cong W^e(\mathbf{S}) \oplus \bar{W}^s(\mathbf{S}) \oplus \bar{W}^u(\mathbf{S})$, $\dim W^e(\mathbf{S}) = 2$, $\dim \bar{W}^s(\mathbf{S}) = \dim \bar{W}^u(\mathbf{S}) = 2$.

Proof: The ordinary cases satisfy 3 conditions, i.e. non-resonant, non-degenerate real saddle, non-degenerate equilibrium. There are 6 ordinary cases, for these cases, eigenvalues are different and non-zero for these ordinary cases.

Resonant cases have at least two purely imaginary eigenvalues coincident, degenerate equilibrium cases have at least two eigenvalues equal to zero, and degenerate real saddle cases have at least two real eigenvalues coincident. So purely resonant equilibrium points have 3 cases, purely degenerate equilibrium points have 7 cases, and purely degenerate real saddle equilibrium points have 3 cases.

For the mixed cases, there exist 1 degenerate-equilibrium and resonant case, as well as 1 degenerate-equilibrium and degenerate real saddle case. There is no resonant and degenerate real saddle case because the resonant case and the degenerate real saddle case need at least 4 eigenvalues on the imaginary axis, respectively. This implies that the resonant and degenerate real saddle cases need at least 8 eigenvalues, but there is only 6 eigenvalues.

Here we proof the conclusion for Case 2 about the structure of the submanifold, the proof for other cases are similar to that for Case 2. In the Case 2, the forms of the eigenvalues are $\pm\alpha_j \left(\alpha_j \in \mathrm{R}, \alpha_j > 0, j = 1\right)$ and $\pm i\beta_j \left(\beta_j \in \mathrm{R}, \beta_j > 0; j = 1, 2 | \beta_1 \neq \beta_2\right)$; the non-zero real eigenvalues $\alpha_j$ correspond to the asymptotically unstable manifold $\bar{W}^u(\mathbf{S})$ while the non-zero real eigenvalues $-\alpha_j$ correspond to the asymptotically stable manifold $\bar{W}^s(\mathbf{S})$, and the purely imaginary eigenvalues $\pm i\beta_j$ correspond to the central manifold $W^c(\mathbf{S})$. The dimension of $\bar{W}^u(\mathbf{S})$, $\bar{W}^s(\mathbf{S})$, and $W^c(\mathbf{S})$ are 1, 1, 4, respectively. There is no resonant manifold in this case. So the structure of the submanifold is $(\mathbf{S}, \Omega) \simeq T\Xi \cong \bar{W}^s(\mathbf{S}) \oplus W^c(\mathbf{S}) \oplus \bar{W}^u(\mathbf{S})$, $\dim W^c(\mathbf{S}) = 4$, $\dim \bar{W}^s(\mathbf{S}) = \dim \bar{W}^u(\mathbf{S}) = 1$, and $\dim W^r(\mathbf{S}) = 0$. □

The classification of the equilibrium points is based on the stability and meticulous dynamical behaviours around equilibrium points. Figure 1 shows the topological classification of six eigenvalues on the complex plane while figure 1 shows classifications and properties of the equilibrium points. Using the conclusion in Section 3.3, one can know the number of periodic orbit families around equilibrium points. For example, if the equilibrium point belong to Case 01, $\dim E^r(L) = 0$, $\dim E^c(L) = 6$, the estimate formula $\frac{1}{2}\left[\dim E^c(L) - \dim E^r(L)\right] + \mathrm{sgn}\left[\dim E^r(L)\right]$ $= 3$, so there are at least 3 families of periodic orbits for the equilibrium point. If the equilibrium point belong to Case R2, $\dim E^r(L) = 4$, $\dim E^c(L) = 6$, the formula equal 2, so there are at least 2 families of periodic orbits for the equilibrium point. If the equilibrium point belong to Case R3, $\dim E^r(L) = 4$, $\dim E^c(L) = 4$, then there are at least 1 family of periodic orbits for the equilibrium point.

Phase diagram on the asymptotically stable manifold $\bar{W}^s(\mathbf{S})$ is the motion approach to the equilibrium point as a sink, while on the asymptotically stable manifold $\tilde{W}^s(\mathbf{S})$ is the motion approach to the equilibrium point as a spiral sink. Phase diagram on the asymptotically unstable manifold $\bar{W}^u(\mathbf{S})$ is the motion leave the equilibrium point as a source, while on the asymptotically unstable manifold $\tilde{W}^u(\mathbf{S})$ is the motion leave the equilibrium point as a spiral source. Phase diagram on the central manifold $W^c(\mathbf{S})$ is the motion around the equilibrium point as a center, and on the resonant manifold $W^r(\mathbf{S})$ is the resonant motion around the equilibrium point. Moreover, phase diagram on the manifold $\bar{W}^s(\mathbf{S})\oplus\bar{W}^u(\mathbf{S})$ is the motion around the equilibrium point as a saddle, while on the manifold $\tilde{W}^s(\mathbf{S})\oplus\tilde{W}^u(\mathbf{S})$ is the motion around the equilibrium point as a spiral saddle.

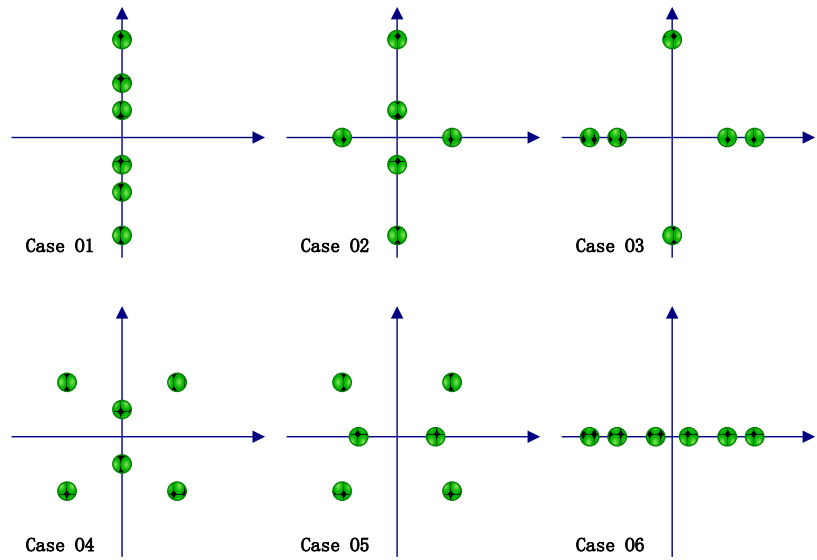

Fig.1a The topological classification of six eigenvalues on the complex plane——ordinary cases

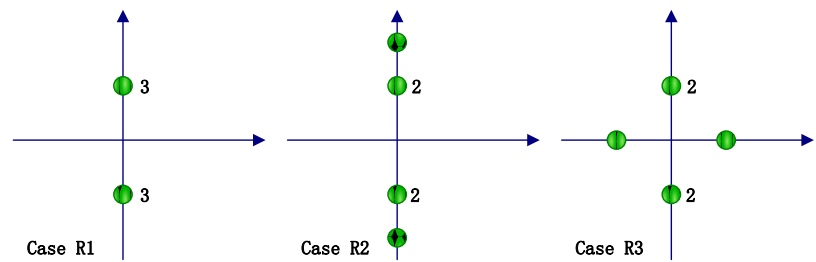

Fig.1b The topological classification of six eigenvalues on the complex plane——resonant cases

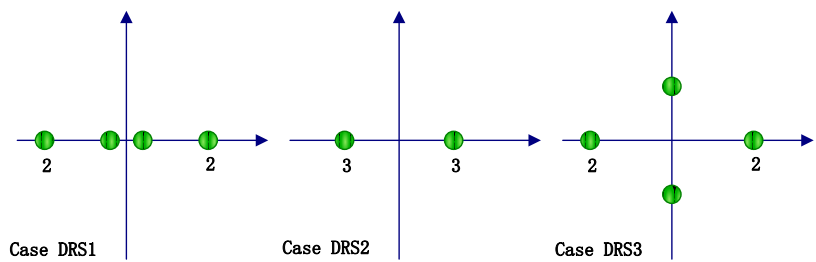

Fig.1c The topological classification of six eigenvalues on the complex plane——degenerate real saddle cases

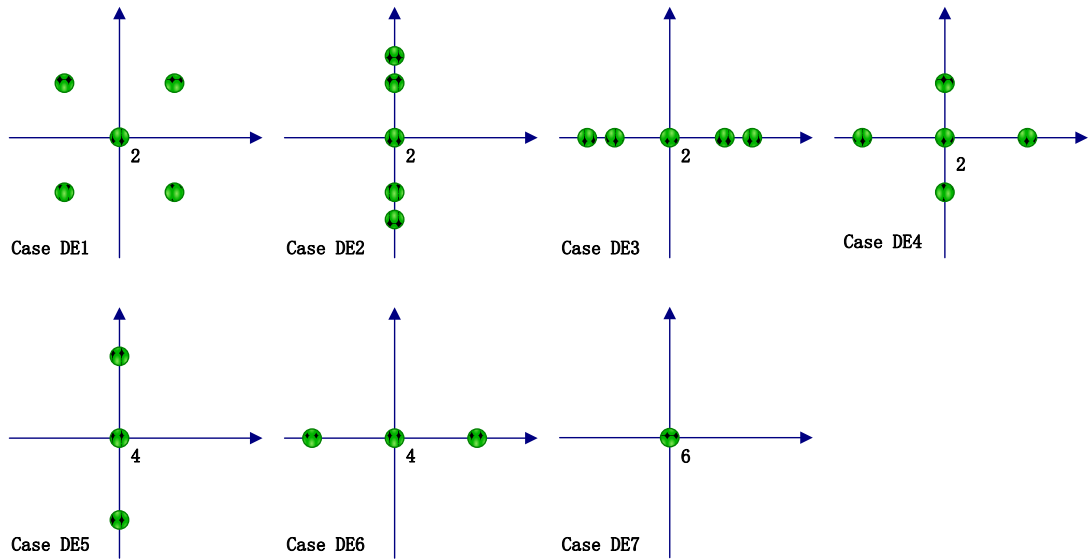

Fig.1d The topological classification of six eigenvalues on the complex plane——degenerate equilibrium cases

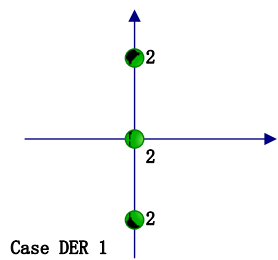

Fig.1e The topological classification of six eigenvalues on the complex plane——degenerate-equilibrium and resonant case

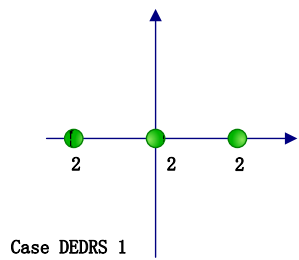

Fig.1f The topological classification of six eigenvalues on the complex plane——degenerate-equilibrium and degenerate real saddle case

# 5. Chaos in the Neighborhood of the Equilibrium Point when Parameter changes

To our opinion, chaos is a sophisticated phenomenon, very tricky and not easy to define or identify. For a dynamical system, the scenarios of the transition to chaos caused by bifurcation sequences with the parameter moving along a relevant direction, and a set of bifurcations resulting in the appearance of a chaotic attractor can be observed [48]. In this section, we will show chaotic motion in the neighborhood of the equilibrium point when some system parameter changes, which includes the sensitive to initial conditions of resonant equilibrium points, the movement of eigenvalues around the resonant case, bifurcation of the motion near equilibrium points as well as the appearing and disappearing of the periodic orbit families in the vicinity of equilibrium points. When the parameters are valued such that the topological cases of equilibrium points is near the resonant cases, a slight change of initial conditions may cause a great change of the final status and the stability of equilibrium points. Periodic orbits in the vicinity of resonant equilibrium points are dense. Besides, near the resonant equilibrium point, the system Eq. (1) with parametric variation is sensitive to initial conditions and topologically mixing. Thus, in the vicinity of resonant equilibrium points, the dynamical system is chaotic.

## 5.1 Sensitive to Initial Conditions of Resonant Equilibrium Points and Bifurcations

Theorem 3 says that if the non-degenerate equilibrium point $\mathbf{r}_0 = \boldsymbol{\tau}(\boldsymbol{\mu}_0) = \boldsymbol{\tau}_0$ in the gravitational field of a rotating body is non-resonant, then, the stability in a sufficiently small open neighborhood of $\boldsymbol{\mu}_0$ is the same as the stability at the point $\boldsymbol{\mu}_0$. The conclusion about the non-degenerate resonant equilibrium point is opposite to the conclusion about the non-degenerate non-resonant equilibrium point. On the basis of Theorem 4, the non-degenerate resonant equilibrium point is a Hopf branching point.

**Corollary 1.** Suppose the non-degenerate equilibrium point $\mathbf{r}_0 = \boldsymbol{\tau}(\boldsymbol{\mu}_0) = \boldsymbol{\tau}_0$ in the gravitational field of a rotating body is a resonant equilibrium point. Denote $(\mathbf{S}, \Omega) \simeq T\Xi \simeq W^c(\mathbf{S}) \oplus Q^c(\mathbf{S})$, where $Q^c(\mathbf{S}) = (\mathbf{S}, \Omega)/W^c(\mathbf{S})$ is the quotient manifold of the central manifold $W^c(\mathbf{S})$, and $W^r(\mathbf{S}) \subseteq W^c(\mathbf{S})$. Then the equilibrium point is a Hopf branching point.

**Proof:** According to Theorem 1, one can know that the non-degenerate equilibrium point is continuous in the presence of persistently acting parameter variation. Then the non-degenerate equilibrium point is exist and continuous when the parameter changes. Consider the three resonant cases as well as the only one degenerate-equilibrium and resonant case, near the equilibrium point, different kinds of purely imaginary eigenvalues appear, which leads to the appearance of new periodic orbits and tori. Thus the equilibrium point is a Hopf branching point.□

**Remark 4**: Consider an equilibrium point $\boldsymbol{\tau}(\boldsymbol{\mu}_0)$, if: for any sufficiently small open neighborhood $G_N(\boldsymbol{\mu}_0)$ of $\boldsymbol{\mu}_0$. $\exists \boldsymbol{\mu}_1, \boldsymbol{\mu}_2 \in G_N(\boldsymbol{\mu}_0)$, such that the equilibrium point $\boldsymbol{\tau}(\boldsymbol{\mu}_1)$ is linear stable restricted to the central manifold $W^c(\mathbf{S})$, and the equilibrium point $\boldsymbol{\tau}(\boldsymbol{\mu}_2)$ is non- resonant and unstable restricted to the central manifold $W^c(\mathbf{S})$. In other words, there is a function $\mathbf{r} = \boldsymbol{\tau}(\boldsymbol{\mu})$ such that $\boldsymbol{\tau}(\boldsymbol{\mu}_0) = \boldsymbol{\tau}_0$, $\mathbf{F}(\boldsymbol{\mu}, \boldsymbol{\tau}) = 0$ for any $\boldsymbol{\mu} \in G_N(\boldsymbol{\mu}_0)$, and $\boldsymbol{\tau}(\boldsymbol{\mu})$ is an equilibrium point for any $\boldsymbol{\mu} \in G_N(\boldsymbol{\mu}_0)$; besides, $\boldsymbol{\tau}(\boldsymbol{\mu}_1)$ is linear stable while $\boldsymbol{\tau}(\boldsymbol{\mu}_2)$ is non-resonant and unstable restricted to the central manifold $W^c(\mathbf{S})$. Then the equilibrium point $\boldsymbol{\tau}(\boldsymbol{\mu}_0)$ is a Hopf branching point.

Thus, we know that the topological structure and dynamical behaviour near resonant equilibrium points are sensitive to initial conditions. For resonant equilibrium points, the resonant cases can change to other cases when eigenvalues moves. Figure 2 shows the movement of eigenvalues around the resonant case R2 which leads to Case O1 $\to$ Case R2 $\to$ Case O4 and Case O4 $\to$ Case R2 $\to$ Case O1, while figure 3 shows the movement of eigenvalues around the degenerate real saddle case DRS3 which leads to Case O4 $\to$ Case DRS3 $\to$ Case O3 and Case O3 $\to$ Case DRS3 $\to$ Case O4. Figure 4 shows the movement of eigenvalues which leads to Case DE7 $\to$ Case O4 and Case O4 $\to$ Case DE7. Figure 5 shows The movement of eigenvalues which leads to Case O2 $\to$ Case DE4 $\to$ Case O3 and Case O3 $\to$ Case DE4 $\to$ Case O2.

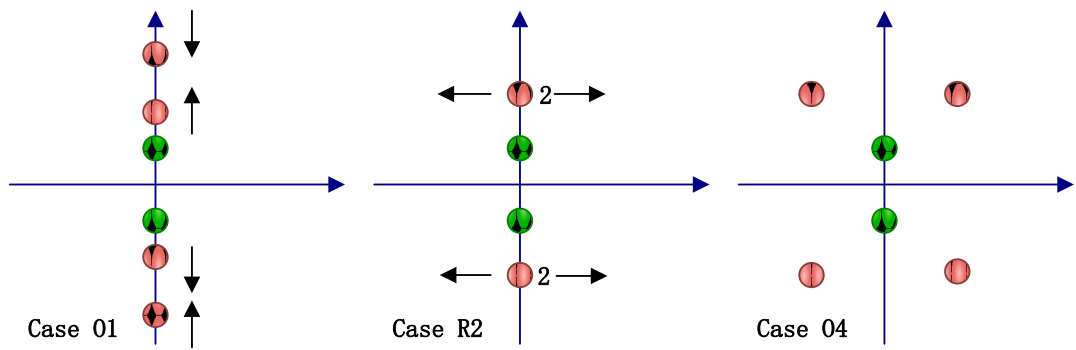

Fig.2a The movement of eigenvalues around the resonant case R2 which leads to
Case O1 $\to$ Case R2 $\to$ Case O4

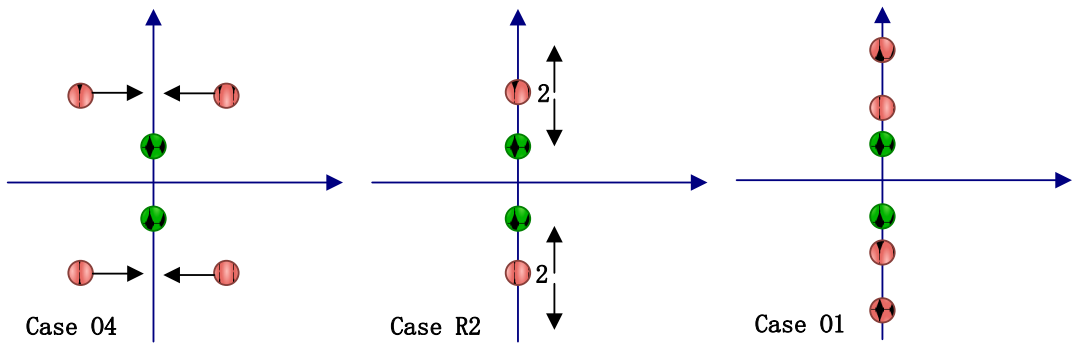

Fig.2b The movement of eigenvalues around the resonant case R2 which leads to
Case O4 → Case R2 → Case O1

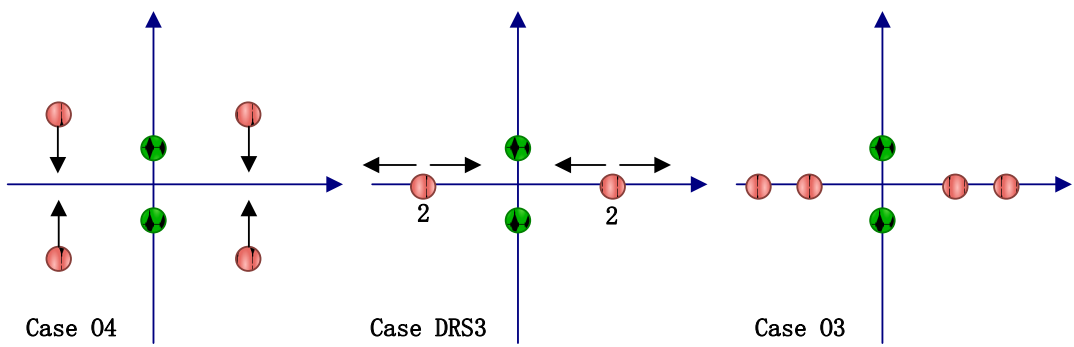

Fig.3a The movement of eigenvalues around the degenerate real saddle case DRS3 which leads to Case O4 → Case DRS3 → Case O3

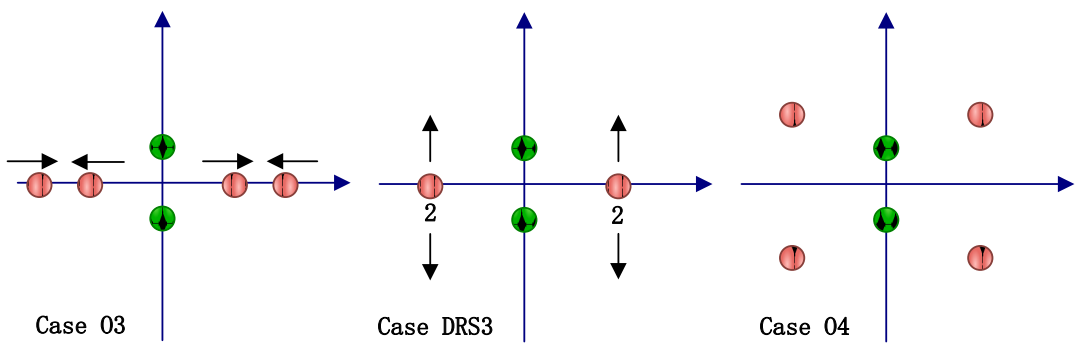

Fig.3b The movement of eigenvalues around the degenerate real saddle case DRS3 which leads to Case O3 → Case DRS3 → Case O4

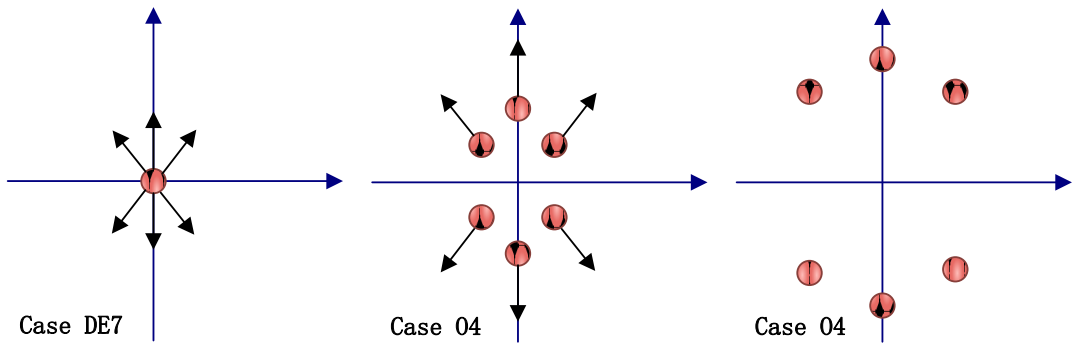

Fig.4a The movement of eigenvalues which leads to Case DE7 → Case O4

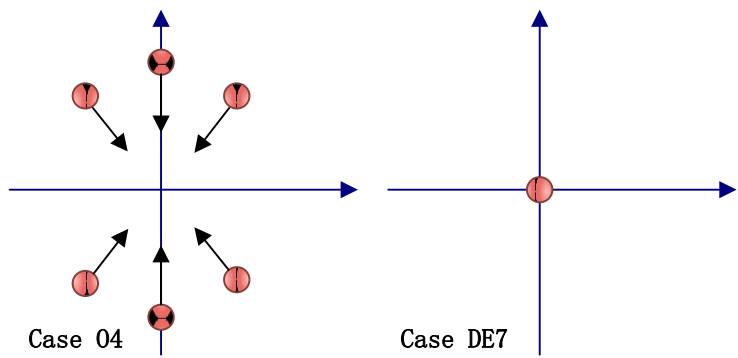

Fig.4b The movement of eigenvalues which leads to Case O4 → Case DE7

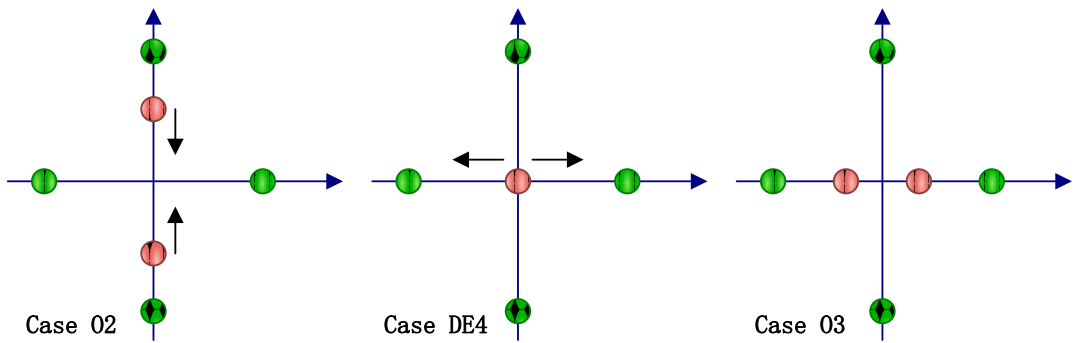

Fig.5a The movement of eigenvalues which leads to Case O2 → Case DE4 → Case O3

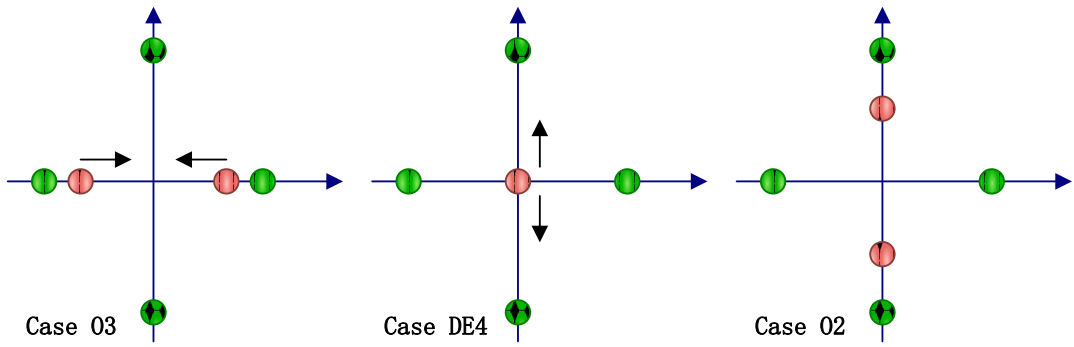

Fig.5b The movement of eigenvalues which leads to Case O3 $\rightarrow$ Case DE4 $\rightarrow$ Case O2

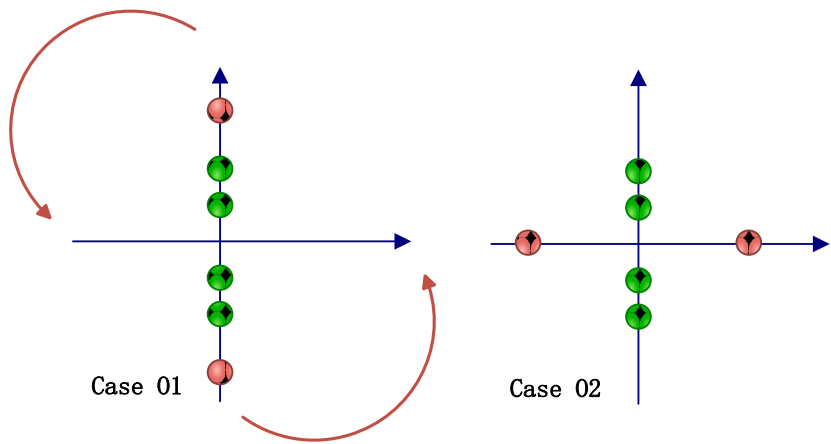

Fig.6 The movement of eigenvalues $+\beta i \rightarrow \infty \rightarrow +\alpha\,(\alpha, \beta \in R)$ which leads to

Case O1 $\rightarrow$ Case O2

## 5.2 Hopf Bifurcations

Before discussing the Hopf bifurcation of the motion near the non-degenerate equilibrium points, a lemma about the eigenvalues is given.

**Lemma 1.** For a non-degenerate equilibrium point, the pure imaginary simple eigenvalues will not leave the imaginary axis while the parameter **μ** has a sufficiently small change. In other words, if a pair of pure imaginary $\pm i\beta_1\,(\beta_1 \in R^+)$ (which are calculated when the parameter equals to $\mathbf{\mu}_0$) satisfy:

there is not an imaginary eigenvalue $i\beta_2 \left(\beta_2 \in \mathbb{R}^+\right)$ satisfy $\beta_1 = \beta_2$, then the pure imaginary $\pm i\beta_1 \left(\beta_1 \in \mathbb{R}^+\right)$ will stay on the imaginary axis for any $\boldsymbol{\mu} \in G_N(\boldsymbol{\mu}_0)$, where $G_N(\boldsymbol{\mu}_0)$ is a sufficiently small open neighborhood of $\boldsymbol{\mu}_0$.

**Proof.** Through the fundamental theorem of algebra, we know that every single-variable, degree $n$ polynomial with complex coefficients has, counted with multiplicity, exactly $n$ roots. Consider the pair of pure imaginary $\lambda_{1,2}(\boldsymbol{\mu}_0) = \pm i\beta_1 \left(\beta_1 \in \mathbb{R}^+\right)$ that corresponds to the parameter $\boldsymbol{\mu}_0$. Suppose that for any sufficiently small open neighborhood of $\boldsymbol{\mu}_0$, which is denoted as $G_N(\boldsymbol{\mu}_0)$, there is a parameter $\boldsymbol{\mu}_1$, such that the eigenvalues $\lambda_{1,2}(\boldsymbol{\mu}_1)$ are not on the imaginary axis. This result illustrate that the sextic characteristic equation has 8 roots, which is contradictory to the fundamental theorem of algebra.□

Lemma 1 means that pure imaginary eigenvalues of the non-degenerate equilibrium point will not leave the imaginary axis before knocked together. In like manner, real eigenvalues of the non-degenerate equilibrium point will not leave the real axis before knocked together.

Consider the manifold $H = h$ near the equilibrium point, where $h = H(L) + \varepsilon^2$, and $\varepsilon^2$ is small enough so that there doesn't exist another equilibrium point $\tilde{L}$ in the small enough open neighborhood on the manifold $(\mathbf{S}, \Omega)$ with the Jacobi constant $H(\tilde{L})$ satisfies $H(L) \leq H(\tilde{L}) \leq h$. The period of the periodic orbit on the manifold $H = h$ is $T(h,\omega) = T(H(L) + \varepsilon^2, \omega)$, it is the function of $h$, and satisfies $\lim_{\varepsilon \to 0} T(h,\omega) \triangleq T(H(L),\omega) = \dfrac{2\pi}{\omega}$, where $\pm i\omega (\omega > 0)$ are eigenvalues at the equilibrium point. A pair of pure imaginary eigenvalues generates one family of periodic orbits.

**Definition 2.** The *characteristic period* for the families of periodic orbits near the non-degenerate equilibrium point is $T(H(L)) = \lim_{\varepsilon \to 0} T(h) = \frac{2\pi}{\omega}$.

It is found that resonant equilibrium points in the potential field of a rotating body are Hopf bifurcation points. The first class of Hopf bifurcation is the equilibrium points belong to Case R1, the forms of the eigenvalues are $\pm i\beta_j (\beta_j \in \mathrm{R}, \beta_1 = \beta_2 = \beta_3 > 0; j = 1, 2, 3)$; then, the structure of the submanifold is $(\mathbf{S}, \Omega) \simeq T\Xi \simeq W^c(\mathbf{S}) \simeq W^r(\mathbf{S})$, and $\dim W^r(\mathbf{S}) = \dim W^c(\mathbf{S}) = 6$. The second class of Hopf bifurcation is the equilibrium points belong to Case R2, the forms of the eigenvalues are $\pm i\beta_j (\beta_j \in \mathrm{R}, \beta_j > 0, \beta_1 = \beta_2 \neq \beta_3; j = 1, 2, 3)$; then, the structure of the submanifold is $(\mathbf{S}, \Omega) \simeq T\Xi \cong W^c(\mathbf{S})$, and $\dim W^r(\mathbf{S}) = 4$. The third class of Hopf bifurcation is the equilibrium points belong to Case R3, the forms of the eigenvalues are $\pm \alpha_j (\alpha_j \in \mathrm{R}, \alpha_j > 0, j = 1), \pm i\beta_j (\beta_j \in \mathrm{R}, \beta_1 = \beta_2 > 0; j = 1, 2)$; then, the structure of the submanifold is $(\mathbf{S}, \Omega) \simeq T\Xi \cong W^s(\mathbf{S}) \oplus W^c(\mathbf{S}) \oplus W^u(\mathbf{S})$, $\dim W^r(\mathbf{S}) = \dim W^c(\mathbf{S}) = 4$, and $\dim W^s(\mathbf{S}) = \dim W^u(\mathbf{S}) = 1$. Theorems 5-7 describe the Hopf bifurcation of the motion near the equilibrium points.

**Theorem 5.** If an equilibrium point $\mathbf{r}_0 = \boldsymbol{\tau}(\boldsymbol{\mu}_0) = \boldsymbol{\tau}_0$ belongs to Case R1, then:

a) There is a sufficiently small open neighborhood of $\boldsymbol{\mu}_0$, which is denoted as $G_N(\boldsymbol{\mu}_0)$, such that for any $\boldsymbol{\mu}_1 \in G_N(\boldsymbol{\mu}_0)$, the equilibrium point $\boldsymbol{\tau}(\boldsymbol{\mu}_1)$ does not belong to Case O2, O3, O5 and O6 among ordinary cases;

b) For any sufficiently small open neighborhood of $\boldsymbol{\mu}_0$, which is denoted as $G_N(\boldsymbol{\mu}_0)$, there is $\boldsymbol{\mu}_1 \in G_N(\boldsymbol{\mu}_0)$, such that the equilibrium point $\boldsymbol{\tau}(\boldsymbol{\mu}_1)$ belongs to one of Case O1, O4, or R2.

**Theorem 6.** If an equilibrium point $\mathbf{r}_0 = \boldsymbol{\tau}(\boldsymbol{\mu}_0) = \boldsymbol{\tau}_0$ belongs to Case R2, then

a) There is a sufficiently small open neighborhood of $\boldsymbol{\mu}_0$, which is denoted as $G_N(\boldsymbol{\mu}_0)$, such that for any $\boldsymbol{\mu}_1 \in G_N(\boldsymbol{\mu}_0)$, the equilibrium point $\boldsymbol{\tau}(\boldsymbol{\mu}_1)$ does not belong to Case O2, O3, O5 and O6 among ordinary cases;

b) For any sufficiently small open neighborhood of $\boldsymbol{\mu}_0$, which is denoted as $G_N(\boldsymbol{\mu}_0)$, there is $\boldsymbol{\mu}_1 \in G_N(\boldsymbol{\mu}_0)$, such that the equilibrium point $\boldsymbol{\tau}(\boldsymbol{\mu}_1)$ belongs to one of Case O1, O4, or R1.

**Theorem 7.** If an equilibrium point $\mathbf{r}_0 = \boldsymbol{\tau}(\boldsymbol{\mu}_0) = \boldsymbol{\tau}_0$ belongs to Case R3, then

a) There is a sufficiently small open neighborhood of $\boldsymbol{\mu}_0$, which is denoted as $G_N(\boldsymbol{\mu}_0)$, such that for any $\boldsymbol{\mu}_1 \in G_N(\boldsymbol{\mu}_0)$, the equilibrium point $\boldsymbol{\tau}(\boldsymbol{\mu}_1)$ does not belong to Case O1, O3, O4, O6 among ordinary cases;

b) For any sufficiently small open neighborhood of $\boldsymbol{\mu}_0$, which is denoted as $G_N(\boldsymbol{\mu}_0)$, there is $\boldsymbol{\mu}_1 \in G_N(\boldsymbol{\mu}_0)$, such that the equilibrium point $\boldsymbol{\tau}(\boldsymbol{\mu}_1)$ belongs to Case O2 or O5.

**Proof for Theorem 5.** From Lemma 1, one can know that for a non-degenerate equilibrium point, the pure imaginary eigenvalues will not leave the imaginary axis before knocked together, and if the imaginary eigenvalues leave the imaginary axis, there are 4 new complex eigenvalues appear, which are in the form of $\pm\sigma \pm i\tau\,(\sigma,\tau \in \mathrm{R}; \sigma,\tau > 0)$, and 4 pure imaginary eigenvalues disappear, which are in the form of $\pm i\beta_1, \pm i\beta_2\,(\beta_1,\beta_2 \in \mathrm{R}^+; \beta_1 = \beta_2)$. So if there is a sufficiently small open neighborhood of $\boldsymbol{\mu}_0$, which is denoted as $G_N(\boldsymbol{\mu}_0)$, such that there is a value $\mu_1 \in G_N(\mu_0)$, the equilibrium point $\tau(\mu_1)$ belongs to Case O2. There are only 2 new complex eigenvalues appear, it contradict the conclusion of appearing of 4 new complex eigenvalues. In other words, there is a sufficiently small open neighborhood $G_N(\mu_0)$, such that for any $\mu_1 \in G_N(\mu_0)$, the equilibrium point $\tau(\mu_1)$ does not belong to Case O2. In like manner, the equilibrium point $\tau(\mu_1)$ does not belong to Case R3. If there is a sufficiently small open neighborhood $G_N(\mu_0)$, such that there is a value $\mu_1 \in G_N(\mu_0)$, the equilibrium point $\tau(\mu_1)$ belongs to Case O5. Then the 2 real eigenvalues for Case O5 will not leave the real axis before equal to zero, and for the non-degenerate equilibrium point, eigenvalues don't equal to zero. This leads to the excluding of Case O5. Consider Case O3. The 4 real eigenvalues for Case O3 will not leave the real axis before knocked together and become to a pair of double eigenvalues, which are in the form of $\pm\alpha_1, \pm\alpha_2 (\alpha_1, \alpha_2 \in R^+; \alpha_1 = \alpha_2)$. This leads to the excluding of Case O3. The above discussion proves the Theorem 5.
□

The proof for Theorem 6 and 7 is similar with the proof for Theorem 5. □

**Corollary 2.** Suppose the equilibrium point $\tau(\mu_1)$ satisfies $\mu_1 \in G_N(\mu_0)$, where $G_N(\mu_0)$ is a sufficiently small open neighborhood of $\mu_0$, then we have

a) If the equilibrium point $r_0 = \tau(\mu_0) = \tau_0$ belongs to Case R1, which means he motion of the spacecraft relative to the equilibrium point is on a 1:1:1 resonant manifold and there is only one family of periodic orbits; then the equilibrium point $\tau(\mu_1)$ belongs to one of the following cases: 1) It is a non-resonant equilibrium point, and there is only one family of periodic orbits; 2) It is a 1:1 resonant equilibrium point, and there are two families of periodic orbits; 3) It is a non-resonant equilibrium point, and there are three families of periodic orbits.

b) If the equilibrium point $\mathbf{r}_0 = \boldsymbol{\tau}(\boldsymbol{\mu}_0) = \boldsymbol{\tau}_0$ belongs to Case R2, which means he motion of the spacecraft relative to the equilibrium point is on a 1:1 resonant manifold and there are two families of periodic orbits; then the equilibrium point $\boldsymbol{\tau}(\boldsymbol{\mu}_1)$ belongs to one of the following cases: 1) It is a non-resonant equilibrium point, and there is only one family of periodic orbits; 2) It is a 1:1:1 resonant equilibrium point, and there is only one family of periodic orbits; 3) It is a non-resonant equilibrium point, and there are three families of periodic orbits.

c) If the equilibrium point $\mathbf{r}_0 = \boldsymbol{\tau}(\boldsymbol{\mu}_0) = \boldsymbol{\tau}_0$ belongs to Case R3, which means he motion of the spacecraft relative to the equilibrium point is on a 1:1 resonant manifold and there is only one family of periodic orbits; then the equilibrium point $\boldsymbol{\tau}(\boldsymbol{\mu}_1)$ belongs to one of the following cases: 1) It is a non-resonant equilibrium point, and there are no periodic orbits; 2) It is a non-resonant equilibrium point, and there are two families of periodic orbits.□

From Corollary 2, one can know that the resonant equilibrium point is a Hopf bifurcation.

## 5.3 The Appearing and Disappearing of the Periodic Orbits and Tori

For the 3 resonant cases R1, R2 and R3, or the 7 degenerate equilibrium cases, or the degenerate-equilibrium and resonant case, or the degenerate-equilibrium and degenerate real saddle case of equilibrium points, sudden change of periodic orbit families or torus families occurs, and the topological structure of submanifold as well as the form of phase diagram are sensitivity to initial values. This leads to the chaos near the equilibrium points. Besides, chaotic motion is not only in the vicinity of equilibrium points, but also extended in the large.

Consider the periodic orbits and tori near the equilibrium point $\boldsymbol{\tau}(\boldsymbol{\mu}_1)$, the appearance and disappearance of the periodic orbits and tori can be found around some kinds of equilibrium points when the parameter $\boldsymbol{\mu}_1 \in G_N(\boldsymbol{\mu}_0)$, where $G_N(\boldsymbol{\mu}_0)$ is a sufficiently small open neighborhood of $\boldsymbol{\mu}_0$, and the equilibrium point $\mathbf{r}_0 = \boldsymbol{\tau}(\boldsymbol{\mu}_0) = \boldsymbol{\tau}_0$ belongs to one of the 3 resonant cases R1, R2 and R3, or one of the 7 degenerate equilibrium cases, or the degenerate-equilibrium and resonant case, or the degenerate-equilibrium and degenerate real saddle case.

**6. Applications to Asteroids**

In this section, the theory developed in the previous sections is applied to asteroids 216 Kleopatra, 2063 Bacchus, and 25143 Itokawa to show the dynamical behavior near a rotating highly irregular-shaped celestial body, including order motion of eigenvalues for the equilibrium points, collision of eigenvalues and chaos near resonant equilibrium points. The physical models of these three asteroids were computed by the radar observation model of Neese [49]. The computational method is the polyhedron method [22,34]. Using this method, the gravitational potential [34] of the asteroid can be calculated by

$$U = \frac{1}{2}G\sigma \sum_{e \in edges} \mathbf{r}_e \cdot \mathbf{E}_e \cdot \mathbf{r}_e \cdot L_e - \frac{1}{2}G\sigma \sum_{f \in faces} \mathbf{r}_f \cdot \mathbf{F}_f \cdot \mathbf{r}_f \cdot \omega_f ,$$

Besides [34],

$$\nabla U = -G\sigma \sum_{e \in edges} \mathbf{E}_e \bullet \mathbf{r}_e \cdot L_e + G\sigma \sum_{f \in faces} \mathbf{F}_f \bullet \mathbf{r}_f \cdot \omega_f ,$$

$$\nabla(\nabla U) = G\sigma \sum_{e \in edges} \mathbf{E}_e \cdot L_e - G\sigma \sum_{f \in faces} \mathbf{F}_f \cdot \omega_f ,$$

where $G=6.67 \times 10^{-11}$ m$^3$kg$^{-1}$s$^{-2}$ is the gravitational constant, $\sigma$ is the density of the body; $L_e$ is the factor of integration, $\omega_f$ is the signed solid angle; all the vectors are body-fixed vectors, $\mathbf{r}_e$ is a vector from the field point to some fixed point on the edge e of face f, $\mathbf{r}_f$ is a vector from the field point to any point in the face plane; $\mathbf{E}_e$ and $\mathbf{F}_f$ are face- and edge-normal vectors, respectively, $\mathbf{E}_e$ and $\mathbf{F}_f$ are the geometric parameters of edges and faces, respectively. In this section, locations of equilibrium points are expressed in the body-fixed frame. The equilibrium points E1-4 of asteroids 216 Kleopatra, 2063 Bacchus, and 25143 Itokawa are defined by the initial positions which are showed in Table 1. The method to calculate the equilibrium points is: a) Use numerical differentiation to calculate $\nabla V = \left( \frac{\partial V(x,y,z)}{\partial x}, \frac{\partial V(x,y,z)}{\partial y}, \frac{\partial V(x,y,z)}{\partial z} \right)$; b) Find a point $(x_0, y_0, z_0)$ and calculate $\nabla V(x_0, y_0, z_0)$; c) Use the Newton method (quasi-Newton method or the method of direction of steepest descent) to find another point $(x_1, y_1, z_1)$, which satisfies $|\nabla V(x_1, y_1, z_1)| < |\nabla V(x_0, y_0, z_0)|$; d) Go several steps in c) when $|\nabla V(x_1, y_1, z_1)| < \varepsilon$, where $\varepsilon$ is a sufficiently small positive number, for example, $\varepsilon = 0.0001$.

Table 1 Initial positions [31,50] of the equilibrium points for the asteroids (rotational angular velocity: $1.0\omega$)

216 Kleopatra

| Equilibrium Points | x (km) | y (km) | z (km) |
| --- | --- | --- | --- |
| E1 | 142.852 | 2.44129 | 1.18154 |
| E2 | -1.16383 | 100.740 | -0.545312 |
| E3 | -144.684 | 5.18829 | -0.272463 |
| E4 | 2.22985 | -102.102 | 0.271694 |

2063 Bacchus

| Equilibrium Points | x (km) | y (km) | z (km) |
|---|---|---|---|
| E1 | 1.14738 | 0.0227972 | -0.000861348 |
| E2 | 0.0314276 | 1.07239 | 0.000711379 |
| E3 | -1.14129 | 0.00806235 | -0.00141486 |
| E4 | 0.0203102 | -1.07409 | 0.000849894 |

25143 Itokawa

| Equilibrium Points | x (km) | y (km) | z (km) |
|---|---|---|---|
| E1 | 0.554478 | -0.00433107 | -0.000061 |
| E2 | -0.0120059 | 0.523829 | -0.000201 |
| E3 | -0.555624 | -0.0103141 | -0.000274 |
| E4 | -0.0158721 | -0.523204 | 0.000246 |

## 6.1 Orderly Movement of Equilibrium Points and Eigenvalues

In this section, the theory developed in the previous sections is applied to the asteroid 216 Kleopatra [51-52] to discuss orderly movement of equilibrium points and eigenvalues acting parameter variation. The estimated bulk density of the asteroid 216 Kleopatra is 3.6 $g \cdot cm^{-3}$ [53], its rotational period is 5.385 h and it reveals a dumbbell-shaped object with overall dimensions of $217 \times 94 \times 81$ km [51-52]. There are two moonlets near 216 Kleopatra, which are Alexhelios (S/2008 (216) 1) and Cleoselene (S/2008 (216) 2) [53].

To study the movement of the equilibrium points and eigenvalues, considering $\mathbf{\mu}_\omega$ and $\mathbf{\mu}_U$ are the parameters for $\mathbf{\omega}$ and $U(\mathbf{r})$, respectively. $\mathbf{\mu} = [\mathbf{\mu}_\omega, \mathbf{\mu}_U]$. Let $\boldsymbol{\mu}_\omega$ varies, and $\boldsymbol{\mu}_U$ keeps invariant, which means that $\boldsymbol{\omega}$ varies and $U(\mathbf{r})$ is invariant. Beside $\boldsymbol{\mu}$ and $V(\boldsymbol{\mu},\mathbf{r})$ vary. Consider a rotating homogeneous cube as an example of simple-shaped body. For the cube, denote $G_g$ as the gravitational constant, $\rho$ as the density, $2a$ as the edge length, Liu et al. (2011) found eight equilibrium points in the potential of a cube with the initial value $G_g = 1, \rho = 1, a = 1,$ and $\omega = 1$. In this paper, let $G_g = 1, \rho = 1, a = 1$, and $\omega$ varies in the interval $[0.5, 2.0]$ with the step-size $0.05$. Figure 7 shows the motion of equilibrium points for a rotating homogeneous cube while the rotational angular velocity changes between $[0.5, 2.0]$ with the step-size $0.05$. The parameter vector $\boldsymbol{\mu}$ follows the same variation. The unique change parameter is rotation rate $\boldsymbol{\omega}$, which comes from the change of $\boldsymbol{\mu}_\omega$ and leads to the change of $\boldsymbol{\mu}$ and $V(\boldsymbol{\mu},\mathbf{r})$. Then the movement of equilibrium points can be seen clearly and the velocity of the movement of equilibrium points leave from the cube faster when the rotational angular velocity of the cube increasing. This provides an example for Theorem 1, which is "The non-degenerate equilibrium point is continuous in the presence of persistently acting parameter variation".

Figure 8 shows the motion of equilibrium points for the asteroid 216 Kleopatra while the rotational angular velocity changes between $[0.5\omega, 2.0\omega]$ with the step-size $0.15\omega$. The parameter vector $\boldsymbol{\mu}$ also follows the same variation. Then the movement of equilibrium points can be seen clearly and the velocity of the movement of equilibrium points leave from the asteroid 216 Kleopatra faster when the rotational angular velocity of the asteroid increasing. Figure 9 shows the motion of eigenvalues affiliated to the equilibrium point E3 for the asteroid 216 Kleopatra while the rotational angular velocity increasing between $[0.5\omega, 2.0\omega]$, the axes labels 'x' and 'y' represent the real part and imaginary part of the eignevalues, respectively. When the rotational angular velocity of the asteroid 216 Kleopatra increasing, eigenvalues leave from the origin point. This is a part of the movement of eigenvalues shown in Figure 4a.

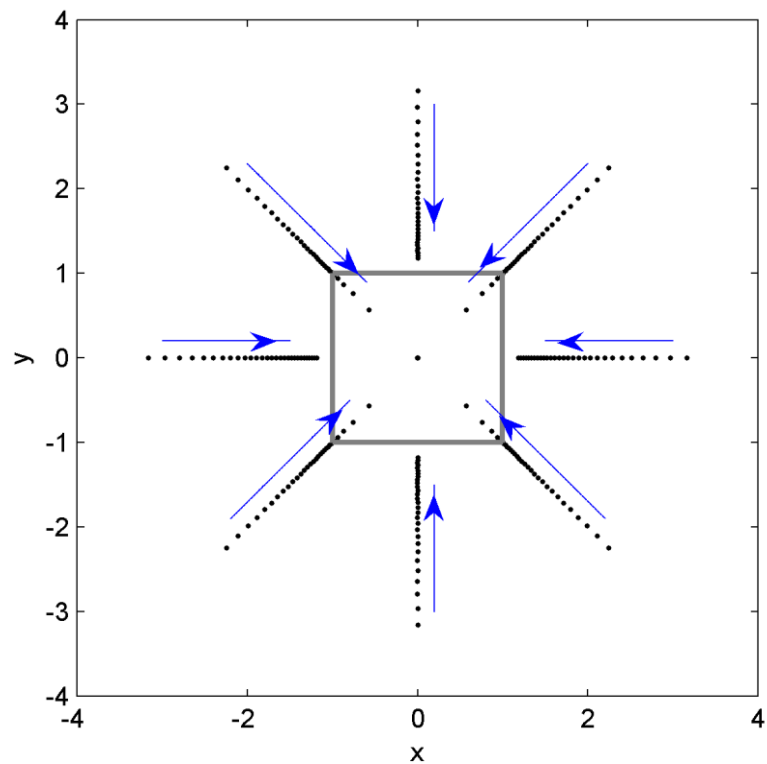

Fig. 7 Motion of equilibrium points for the cube while the rotational angular velocity changes between $[0.5, 2.0]$

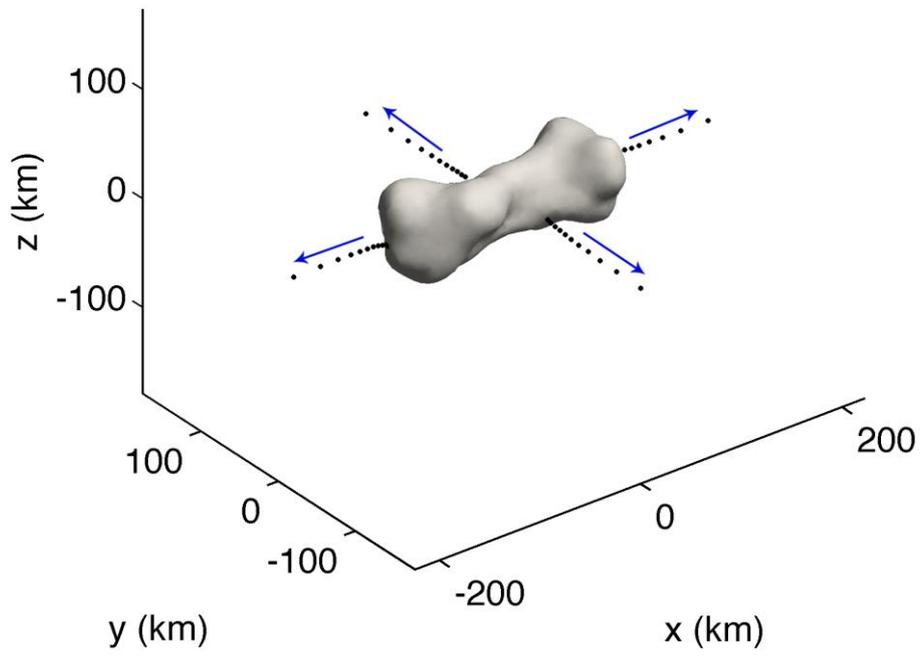

Fig, 8a

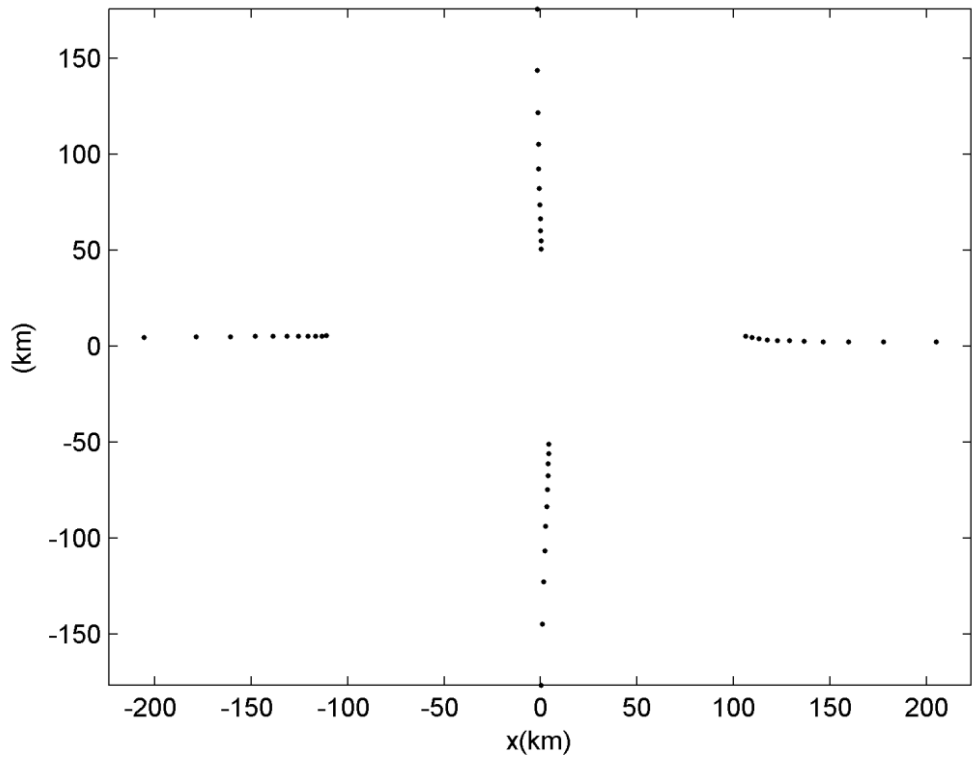

Fig. 8b

Fig. 8 Motion of equilibrium points for the asteroid 216 Kleopatra while the rotational angular velocity changes between $[0.5\omega, 2.0\omega]$

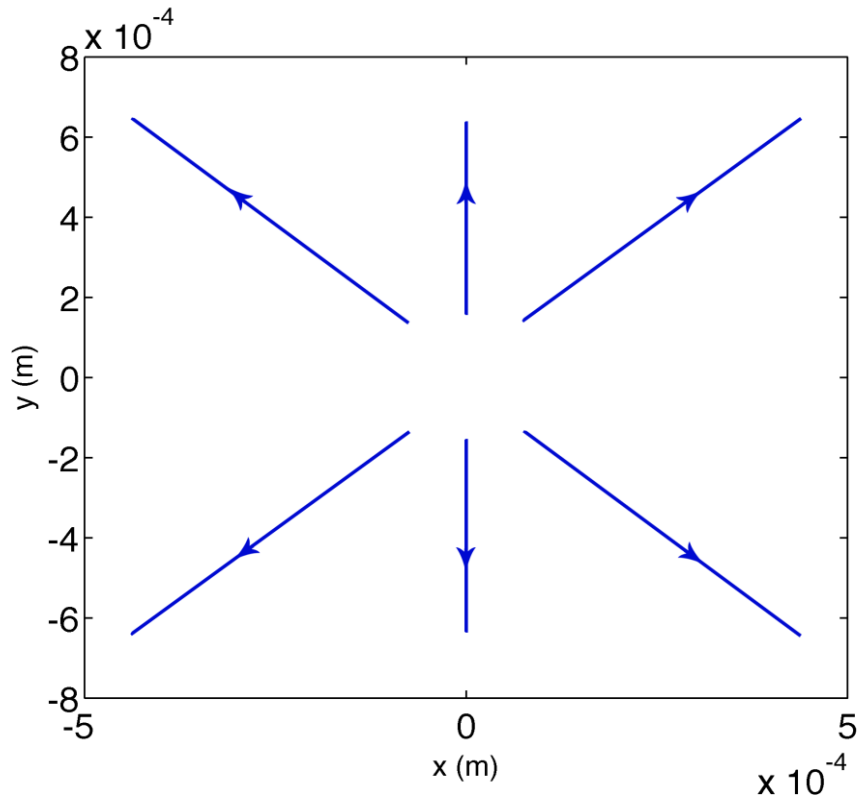

Fig. 9 Motion of eigenvalues affiliated to the equilibrium point E3 for the asteroid 216 Kleopatra while the rotational angular velocity changes between $[0.5\omega, 2.0\omega]$ corresponding to the movement of eigenvalues shown in Fig.5a

## 6.2 Collision of Eigenvalues, Chaos near Resonant Equilibrium Points and Chaos in the Large

In this section, the theory developed in the previous sections is applied to the asteroid 2063 Bacchus [54-55] and 25143 Itokawa [56-57] to discuss the collision of eigenvalues, chaos near resonant equilibrium points as well as chaos of orbits in the large around asteroid 216 Kleopatra. The unique change parameter is also the rotation rate $\boldsymbol{\omega}$, which comes from the change of $\boldsymbol{\mu}_\omega$ and leads to the change of $\boldsymbol{\mu}$ and $V(\mathbf{\mu}, \mathbf{r})$. The rotational period of the asteroid 2063 Bacchus is 14.904 h [54] and the three axis lengths are $1.11 \times 0.53 \times 0.50$ km [55], the density is about $2.0 \text{g cm}^{-3}$ and the total mass is $3.3 \times 10^{12}$ kg [54-55]. The total mass estimate of the asteroid 25143 Itokawa is $4.5^{+2.0}_{-1.8} \times 10^{10}$ kg [56] with the density $1.95 \pm 0.14$ g cm$^{-3}$ [58-59], its rotational period is 12.132 h and the three axis lengths are $535 \times 294 \times 209$ m [60-61].

Figure 10 shows the motion of equilibrium points for the asteroid 2063 Bacchus while the rotational angular velocity changes between $[0.5\omega, 2.0\omega]$ with the step-size $0.075\omega$. The velocity of the movement of equilibrium points also leave from the asteroid 2063 Bacchus faster when the rotational angular velocity of the asteroid is increasing.

Figure 11 shows the motion of eigenvalues affiliated to the equilibrium point E3 for the asteroid 2063 Bacchus while the rotational angular velocity changes between $[0.5\omega, 2.0\omega]$. The axes labels 'x' and 'y' represent the real part and imaginary part of the eignevalues, respectively. In Figure 11, collision of eigenvalues occurs, corresponds to the movement of eigenvalues that leads to Case O1 → Case R2 → Case O4 and is shown in Figure 2a in Sect. 6. When the values of ω equals to 1.56, the purely eigenvalues collide. After collision, the eignevalues will leave the imaginary axis. Eigenvalues before and most close to collision are colored into green, while eigenvalues after and most close to collision are colored into red, others are blue. The second figure in Figure 12 is amplified near the position where eigenvalues collide each other.

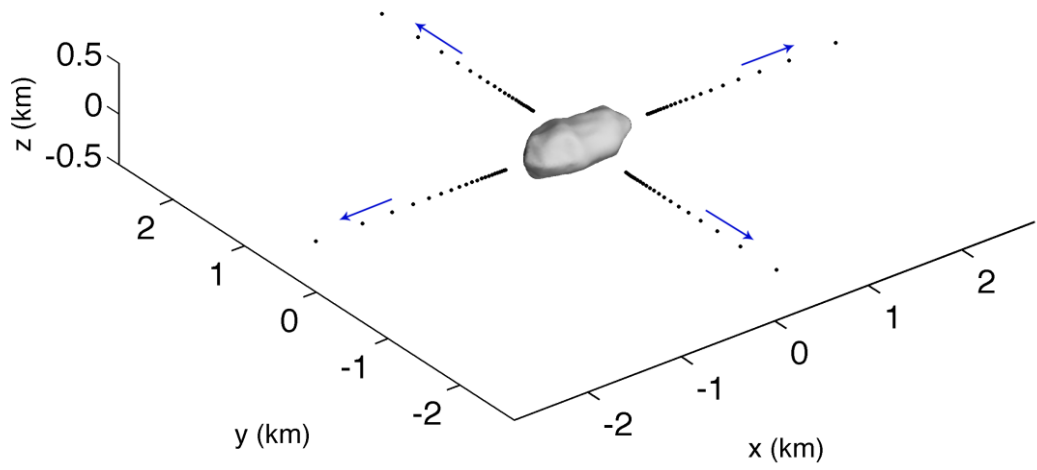

Fig. 10a

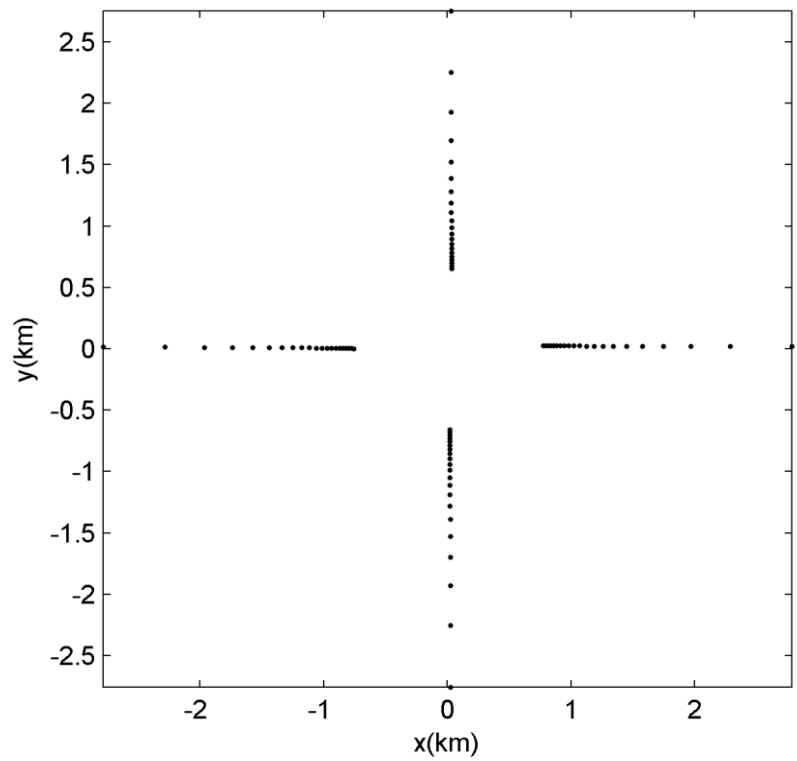

Fig. 10b

Fig. 10 Motion of equilibrium points for the asteroid 2063 Bacchus while the rotational angular velocity changes between $[0.25\omega, 2.0\omega]$

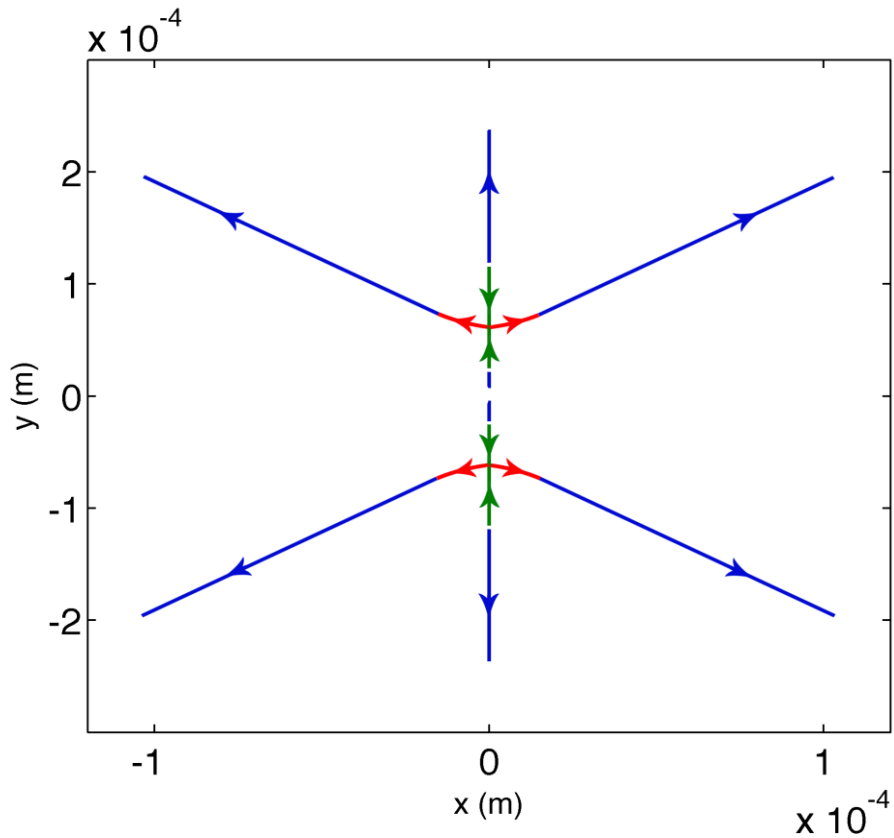

Fig. 11 Motion of eigenvalues affiliated to the equilibrium point E3 for the asteroid 2063 Bacchus while the rotational angular velocity changes between $[0.25\omega, 2.0\omega]$, which corresponds to the movement of eigenvalues that leads to Case O1 $\rightarrow$ Case R2 $\rightarrow$ Case O4 and is shown in Fig. 2a. Green points are eigenvalues before collision while red points are eigenvalues after collision, other eigenvalues are blue. The second figure is amplified near the position with eigenvalues collided.

Figure 12 shows the motion of equilibrium points for the asteroid 25143 Itokawa while the rotational angular velocity changes between $[0.5\omega, 2.0\omega]$ with the step-size $0.075\omega$. The velocity of the movement of equilibrium points also leave from the asteroid 25143 Itokawa faster when the rotational angular velocity of the asteroid increasing.

Figure 13 shows the motion of eigenvalues affiliated to the equilibrium point E3 for the asteroid 25143 Itokawa while the rotational angular velocity changes between $[0.5\omega, 2.0\omega]$. The axes labels 'x' and 'y' represent the real part and imaginary part of the eignevalues, respectively. In Figure 13, collision of eigenvalues occurs, corresponds to the movement of eigenvalues that leads to Case O1 → Case R2 → Case O4 and is shown in Figure 2a in Sect. 6. Eigenvalues before and most close to collision are colored into green, while eigenvalues after and most close to collision are colored into red, others are blue. The second figure in Figure 13 is amplified near the position with eigenvalues collided. One can see that the motion of eigenvalues affiliated to the equilibrium point E3 for the asteroid 25143 Itokawa while the rotational angular velocity changes between $[0.5\omega, 2.0\omega]$ also follows Fig. 13b.

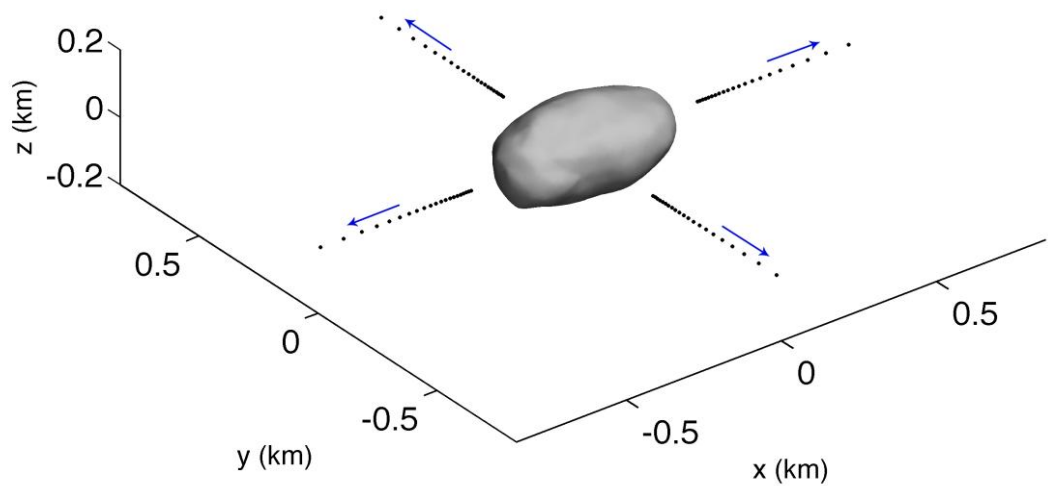

Fig. 12a

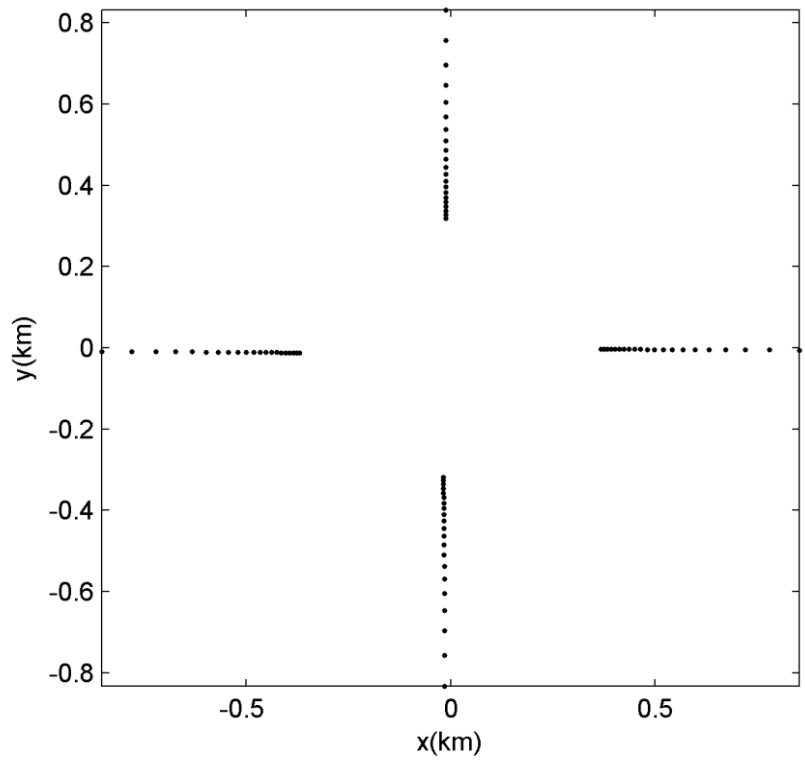

Fig. 12b

Fig. 12 Motion of equilibrium points for the asteroid 25143 Itokawa while the rotational angular velocity changes between $[0.5\omega, 2.0\omega]$

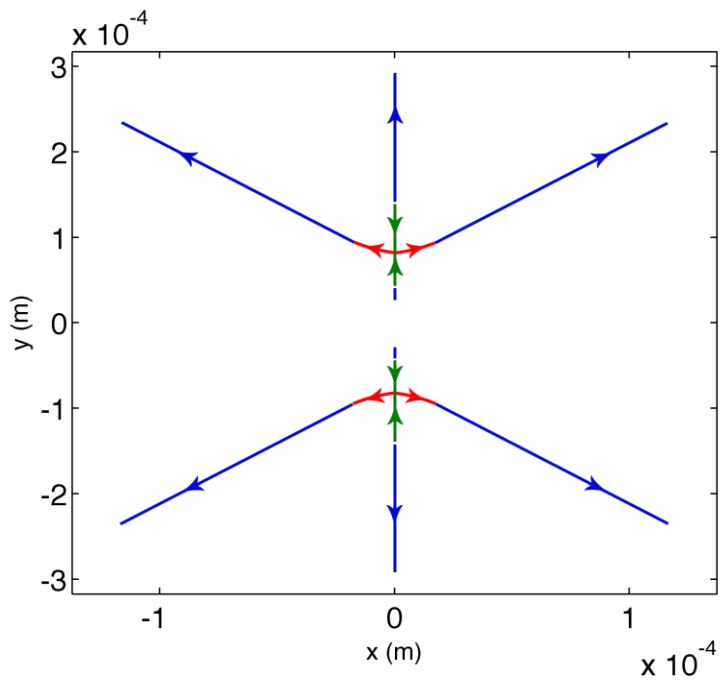

Fig. 13a

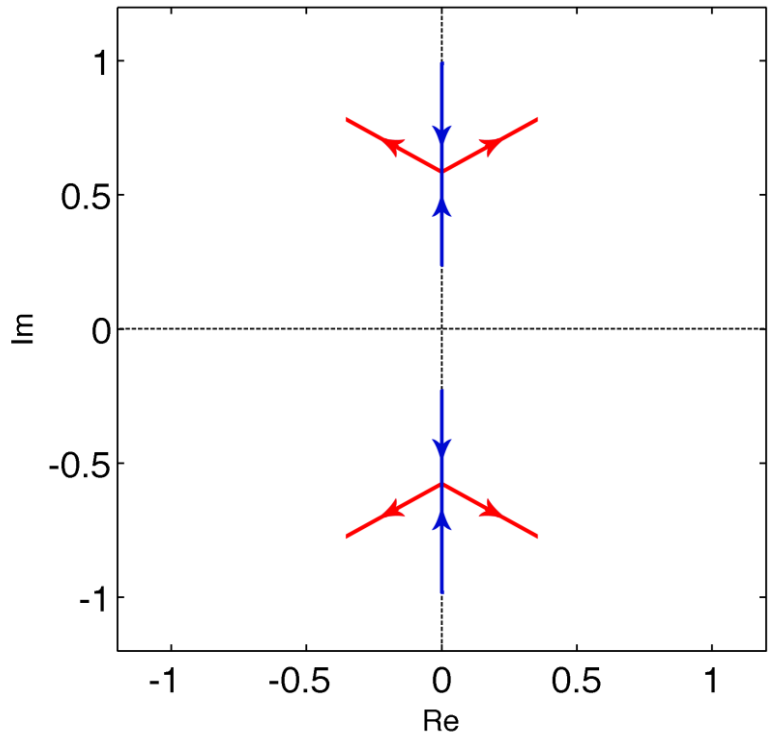

Fig.13b

Fig. 13 Motion of eigenvalues affiliated to the equilibrium point E3 for the asteroid 25143 Itokawa while the rotational angular velocity changes between $[0.5\omega, 2.0\omega]$, which corresponds to the movement of eigenvalues that leads to Case O1→R2→O4 and is shown in Fig. 2a. Green points are eigenvalues before collision while red points are eigenvalues after collision, other eigenvalues are blue. The second figure is amplified near the position with eigenvalues collided.

Not only motion near resonant equilibrium points may leads to chaos, but also orbits in the large can represent chaos behaviors. Two Poincaré surfaces of sections are calculated in the potential of the asteroid 216 Kleopatra to show the chaos behaviors in the large. The dynamical system is integrated by the RK7(8) method and the parameters $\omega$, $\mu$, $\mu_\omega$ and $\mu_U$ are invariant. The orbit is close to the equatorial plane, i.e. the xy plane. The Poincaré surface of sections are defined by $x=0$ and $\dot{x}=0$ in figure 14. Figure 14a shows the Poincaré surface of sections ($x=0$) of orbits in the potential of the asteroid 216 Kleopatra: $y-v_y$, where $v_y=\dot{y}$; while figure 14b shows Poincaré surface of sections ($\dot{x}=0$). The isolated points in Figure 14 represent chaotic orbits. Besides, in figure 14, one can see that among these points in the Poincaré surface of sections, there exist areas with no points; in other words, there is no any orbits reach these areas. From these two Poincaré surfaces of sections, it can be seen that the chaotic motion is complicated and unordered.

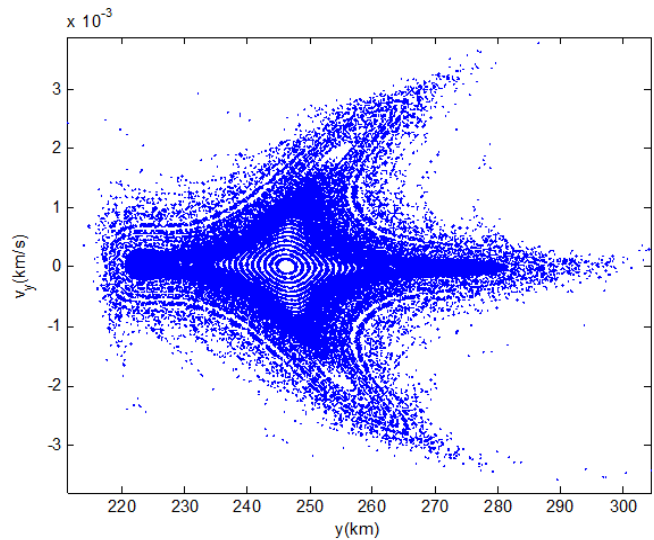

Fig. 14a Poincaré surface of sections ($x=0$) of orbits in the potential of the asteroid 216 Kleopatra: $y-v_y$, where $v_y=\dot{y}$

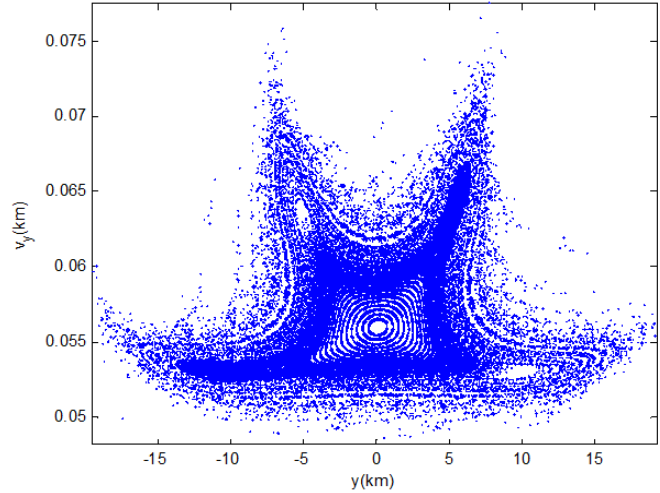

Fig. 14b Poincaré surface of sections ($\dot{x} = 0$) of orbits in the potential of the asteroid 216 Kleopatra: $y - v_y$, where $v_y = \dot{y}$

## 7. Conclusions

This paper is a first step toward the order and chaos of the motion near equilibrium points in the potential of a rotating highly irregular-shaped celestial body. It is discovered that the non-degenerate equilibrium point is continuous in the presence of persistent system parameter variation. Two theorems about the structural stability of non-degenerate and non-resonant equilibrium points are presented and proved. It is found that the topological structure in the vicinity of equilibrium points has 6 ordinary cases, 3 resonant cases, 3 degenerate real saddle cases, 1 degenerate-equilibrium and resonant case, as well as 1 degenerate-equilibrium and degenerate real saddle case.

It is found that resonant equilibrium points are sensitive to initial conditions and non-degenerate resonant equilibrium points are Hopf branching points. Near resonant equilibrium points, one can find new periodic orbits and tori appear, or old periodic orbits and tori disappear. The possible topological transfer between different cases for equilibrium points is systematically given in a figure; besides, eigenvalues can move to the point at infinity.

The theory developed here are suitable for all kinds of rotating celestial bodies, simple-shaped or highly irregular-shaped, including asteroids, comets, planets and satellites of planets. Three highly irregular-shaped asteroids are selected for applications: asteroid 216 Kleopatra, 2063 Bacchus, and 25143 Itokawa. When parameter of the rotational angular velocity changing in some range, the equilibrium points of asteroids 216 Kleopatra and 25143 Itokawa move continuous; eigenvalues affiliated to the equilibrium point E3 for the asteroid 2063 Bacchus and 25143 Itokawa move though the resonant cases, which leads to $\text{Case O1} \rightarrow \text{Case R2} \rightarrow \text{Case O4}$, collision of eigenvalues occurs. The Poincaré sections are calculated in the potential of the asteroid 216 Kleopatra to show the chaos behaviors of large scale orbits, which means that chaos occurs not only locally, but also largely.

## Acknowledgements

This research was supported by the National Basic Research Program of China (973 Program, 2012CB720000), the grant from the State key Laboratory of Astronautic Dynamics (2014-ADL-DW0201), and the National Natural Science Foundation of China (No. 11372150).